\documentclass[sigconf]{acmart}

\usepackage{xspace}

\usepackage{paralist}

\usepackage{hyperref}
\hypersetup{
    colorlinks=true,
    linkcolor=blue,
    filecolor=magenta,      
    urlcolor=cyan,
    citecolor=blue,
    pdfpagemode=FullScreen,
}

\usepackage{pifont}
\usepackage{xcolor,colortbl}
\usepackage{adjustbox}
\usepackage{MnSymbol}%
\usepackage{wasysym}%
\usepackage{subcaption}%

\usepackage{algorithm}
\usepackage{algpseudocode}

\usepackage{tikz}
\usetikzlibrary{shapes, calc, automata}
\usetikzlibrary{positioning}
\usetikzlibrary{overlay-beamer-styles}
\usetikzlibrary{arrows, arrows.meta}
\usetikzlibrary{decorations, decorations.pathreplacing, decorations.markings}
\usetikzlibrary{shapes.multipart} 
\usetikzlibrary{backgrounds}
\usepackage{tikzscale}
\usepackage{multirow}

\usepackage{soul}
\usepackage{soulpos}
\usepackage{contour}
\usepackage[normalem]{ulem} 
\usepackage{expl3}

\definecolor{sangria}{rgb}{0.57, 0.0, 0.04}
\definecolor{pinegreen}{rgb}{0.0, 0.47, 0.44}
\definecolor{rossocorsa}{rgb}{0.83, 0.0, 0.0}
\definecolor{ao}{rgb}{0.0, 0.0, 1.0}
\definecolor{deepjunglegreen}{rgb}{0.0, 0.29, 0.29}
\definecolor{dartmouthgreen}{rgb}{0.05, 0.5, 0.06}

\definecolor{blue1}{HTML}{bee9e8}
\definecolor{blue2}{HTML}{62b6cb}
\definecolor{blue3}{HTML}{cae9ff}
\definecolor{blue4}{HTML}{5fa8d3}
\definecolor{offwhite}{HTML}{D3D3D3} 
\definecolor{lightgray}{HTML}{f3f3f3}
\newcommand{\toolname}{\textsc{Pemu}\xspace}
\newcommand{\eg}{e.g.,\xspace}
\newcommand{\ie}{i.e.,\xspace}
\newcommand{\etal}{et~al.\@\xspace}

\newcommand{\aflnet}{\textsc{AflNet}\xspace}

\newcommand{\emnettest}{\textsc{EmNetTest}\xspace}

\newcommand{\vn}{\emph{virtual network}\xspace}
\newcommand{\semu}{\textsc{SEmu}\xspace}
\newcommand{\semubase}{\textsc{SEmu\_base}\xspace}
\newcommand{\semuvn}{\textsc{SEmu\_\toolname}\xspace}
\newcommand{\fuzzware}{\textsc{Fuzzware}\xspace}
\newcommand{\fuzzwarebase}{\textsc{Fuzzware\_base}\xspace}
\newcommand{\fuzzwarerand}{\textsc{Fuzzware\_rand}\xspace}
\newcommand{\fuzzwarevn}{\textsc{Fuzzware\_\toolname}\xspace}
\newcommand{\hoedur}{\textsc{Hoedur}\xspace}
\newcommand{\hoedurbase}{\textsc{Hoedur\_base}\xspace}
\newcommand{\hoedurrand}{\textsc{Hoedur\_rand}\xspace}
\newcommand{\hoedurvn}{\textsc{Hoedur\_\toolname}\xspace}
\newcommand{\psquaredim}{\textsc{P2IM}\xspace}

\newcommand{\rulesep}{\unskip\ {\textcolor{gray}{\vrule width 0.5pt}}\ }

\newcommand{\packer}{encapsulation module\xspace}
\newcommand{\Packer}{Encapsulation Module\xspace}
\newcommand{\parser}{state extractor\xspace}

\newcommand{\circleone}{\ding{202}\xspace}
\newcommand{\circletwo}{\ding{203}\xspace}

\AtBeginDocument{%
  }

\begin{document}

\title{Technical Report: Protocol-Aware Firmware Rehosting for Effective Fuzzing of Embedded Network Stacks}

\author{Moritz Bley}
\email{moritz.bley@cispa.de}
\affiliation{%
  \institution{CISPA Helmholtz Center for
Information Security}
  \city{Saarbr\"ucken}
  \country{Germany}
}

\author{Tobias Scharnowski}
\email{tobias.scharnowski@cispa.de}
\affiliation{%
  \institution{CISPA Helmholtz Center for
Information Security}
  \city{Saarbr\"ucken}
  \country{Germany}
}

\author{Simon W\"orner}
\email{simon.woerner@cispa.de}
\affiliation{%
  \institution{CISPA Helmholtz Center for
Information Security}
  \city{Saarbr\"ucken}
  \country{Germany}
}

\author{Moritz Schloegel}
\email{moritz.schloegel@asu.edu}
\affiliation{%
  \institution{Arizona State University}
  \city{Tempe}
  \state{AZ}
  \country{USA}
}

\author{Thorsten Holz}
\email{thorsten.holz@mpi-sp.org}
\affiliation{
  \institution{Max Planck Institute for
Security and Privacy}
  \city{Bochum}
  \country{Germany}
}

\begin{abstract}

One of the biggest attack surfaces of embedded systems is their network interfaces, which enable communication with other devices.
Unlike their general-purpose counterparts, embedded systems are designed for specialized use cases, resulting in unique and diverse communication stacks.
Unfortunately, current approaches for evaluating the security of these embedded network stacks require manual effort or access to hardware, and they generally focus only on small parts of the embedded system. 
A promising alternative is \emph{firmware rehosting}, which enables fuzz testing of the entire firmware by generically emulating the physical hardware. However, existing rehosting methods often struggle to meaningfully explore network stacks due to their complex, multi-layered input formats. This limits their ability to uncover deeply nested software faults.

To address this problem, we introduce a novel method to automatically detect and handle the use of network protocols in firmware called \toolname. 
By automatically deducing the available network protocols, \toolname can transparently generate valid network packets that encapsulate fuzzing data, allowing the fuzzing input to flow directly into deeper layers of the firmware logic.
Our approach thus enables a deeper, more targeted, and layer-by-layer analysis of firmware components that were previously difficult or impossible to test. 
Our evaluation demonstrates that \toolname consistently improves the code coverage of three existing rehosting tools for embedded network stacks.
Furthermore, our fuzzer rediscovered several known vulnerabilities and identified five previously unknown software faults, highlighting its effectiveness in uncovering deeply nested bugs in network-exposed code.

\emph{This technical report accompanies our paper ``Protocol-Aware Fimrware Rehosting for Effective Fuzzing of Embedded Network Stacks''~\cite{bley2025pemu} published at ACM CCS 2025, and it features more details, most noteworthy Figures~\ref{fig:semuplots},~\ref{fig:fuzzwareplots}, and~\ref{fig:hoedurplots} in the appendix.}
\end{abstract}

\setcopyright{none} 
\settopmatter{printacmref=false, printfolios=true,printccs=true} 
\renewcommand\footnotetextcopyrightpermission[1]{} 
\pagestyle{plain}

\maketitle

\section{Introduction}

Embedded devices are specialized computers that tightly integrate with the hardware they control. Examples include medical devices, industrial control systems, and Internet of Things (IoT) components. These devices are either powered by lightweight operating systems (OSes) or operate without an OS (referred to as ``bare metal''). Consequently, the software running on these systems, known as \emph{firmware}, cannot rely on conventional OS abstractions but must directly manage its interactions with hardware. Furthermore, embedded devices often expose network-related applications through Embedded Network Stacks (ENS) to allow for interconnectivity with other devices. The combination of missing OS abstractions, limited computation resources, and the large attack surface has made embedded systems challenging to secure, especially via automated techniques.

Recent research has turned towards \emph{firmware rehosting} to achieve scalable and effective testing of embedded firmware. Generally speaking, rehosting runs firmware within an emulator instead of its original, resource-constrained hardware environment~\cite{muench2018wycinwyc,wright_challenges_2021}. Using a fuzzer to mimic peripheral data allows firmware to be executed on powerful hosts without relying on the physical microcontroller~\cite{scharnowski2022fuzzware,scharnowski2023hoedur,feng2020p2im,yun_fuzzing_2022,hernandez2022firmwire, seidel2023safirefuzz}.

However, testing firmware that uses networking via rehosting-based fuzzing still faces two substantial challenges: First, while AFL-like mutations~\cite{afl} are effective in testing lightly structured formats, current solutions are ill-equipped to progress past the initial layers of embedded network stacks. Consequently, for firmware that utilizes a complex ENS, current rehosting-based firmware fuzzers are stuck, as they fail to generate the highly complex structure of network packets. This forces them to re-test only the low-level logic that verifies the packet structure. In particular, they fail to test higher network layers or the application-level logic built on top of the ENS. Second, while network protocol fuzzers exist in the general-purpose domain~\cite{pham2020aflnet,amusuo2023systematically,bars2024fuzztructionnet}, their applicability to embedded firmware is limited.
For example, \aflnet~\cite{pham2020aflnet} relies on existing (target-specific) traffic captures for testing. While such a capture is easy to generate on a general-purpose system, this is not the case in the embedded domain. Aside from various practical challenges, the requirement of a hardware setup to bootstrap the testing process results in a tedious manual process. This effectively negates the scalability effects that rehosting provides.
Similarly, \emnettest~\cite{amusuo2023systematically} also relies on a set of seed network packets, for which it systematically generates variants to test the known layers of an ENS. Although more broadly applicable to embedded network stacks, \emnettest requires source code and knowledge about the target to specify network metadata. Both are rarely available in practice. Furthermore, it does not provide fuzz testing capabilities for the firmware application logic beyond the transport layer.

Based on these observations, we identify two properties of an optimal rehosting solution for effective fuzzing of ENSs: First, neither a physical hardware setup nor target-specific configurations should be required, as both introduce manual effort. More specifically, this means that \emph{inputs need to be bootstrapped without any seeds}, and any \emph{network protocols need to be automatically inferred} from firmware behavior. Second, the desired solution should allow the fuzzer to test \emph{not only common network layers}, but also the \emph{application logic that is specific to each firmware}.

\begin{figure}[tb]
    \centering
    \includegraphics[width=\columnwidth]{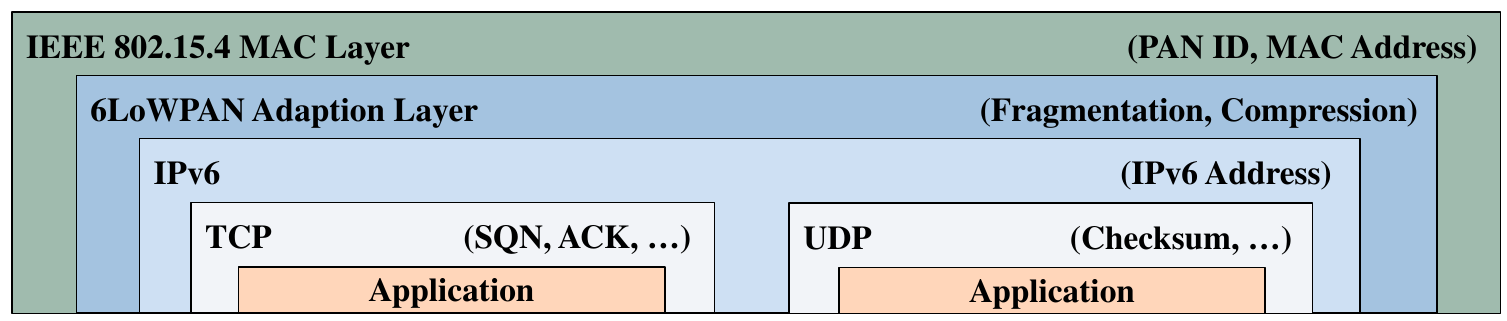}
    \caption{Example network layers involved in a firmware that exposes applications through a typical 6LoWPAN stack.}
    \label{fig:intro-motivation-layer}
\end{figure}

\begin{figure}[tb]
    \centering   
    \includegraphics[width=.78\columnwidth]{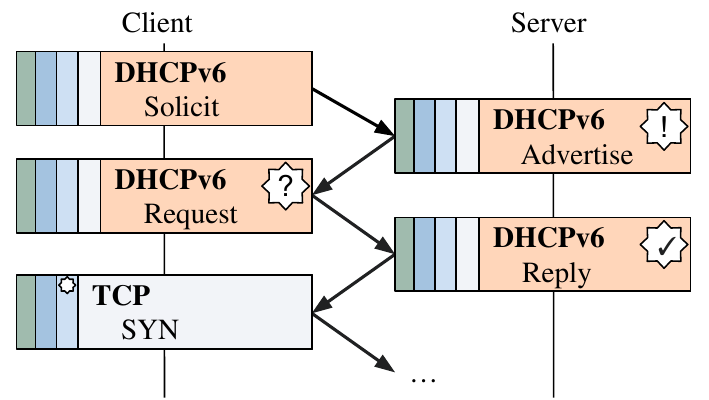}
    \caption{Exchange of context-sensitive messages involved in an IPv6 address retrieval via DHCPv6 between a client (left) and server (right).}%
    \label{fig:intro-motivation-msg-seq}
    \vspace{-.5cm}
\end{figure}

We need to solve different challenges to achieve such automation and testing flexibility. First, we must generate well-formed messages for embedded network protocols. Figure~\ref{fig:intro-motivation-layer} shows a typical ENS that exposes IPv6-based communication through 6LoWPAN over IEEE 802.15.4 radio frames. The firmware expects to receive low-level radio frames, which it decompresses and re-assembles into IPv6 packets. To enable a fuzzer to test the application logic of the firmware under test, we need to wrap application-layer data into a series of low-level radio frames that successfully traverse all network layers and deliver the fuzzing payload to the application logic. This is in contrast to typical general-purpose network fuzzing scenarios, where the operating system abstracts away most network layers via network sockets..
Given that we do not have any information on the ENS of the firmware, we need a mechanism that allows us to automatically discover the tree of different network protocols via which the ENS expects to communicate. Additionally, we also need to infer the static device identifiers by which the firmware expects to be addressed, such as the PAN ID, as well as the MAC and IPv6 addresses. After deriving this information, we can focus on generating valid network packets that encapsulate fuzzing inputs. 
However, to fully explore the ENS, the fuzzer also needs to be able to account for the dynamic state of protocols. For example, Figure~\ref{fig:intro-motivation-msg-seq} displays how the DHCPv6 protocol relies on a defined sequence of messages, including DHCPv6 solicit, advertise, request, and reply packets that must occur in a specific order. Only if the rehosting environment correctly handles these exchanges can the fuzzer fully explore the protocol implementation and any application built on top of it.

To this end, we design and implement \toolname, a generic framework that extends existing rehosting platforms to handle network communications.
At a high level, our approach provides a self-configuring virtual network interface that connects a network-unaware fuzzer and the firmware under test. \toolname encapsulates raw fuzzing input into valid network packets, enabling fuzzing input to reach deeper into network-related code, all the way into the application logic.
To infer the network protocols the firmware under test uses, \toolname performs active probing.
We apply two techniques to monitor how the firmware responds to different types of probing packets. First, in contrast to traditional rehosting systems that largely ignore firmware output, we parse the outgoing low-level frames of the firmware to detect available network protocols. Second, we evaluate the firmware coverage with different types of probing packets. This allows us to detect which types of packets trigger unique coverage in the firmware, \ie which types of packets its ENS is sensitive to. By interleaving fuzzing and active probing based on previously detected network layers, \toolname iteratively recovers the tree of network protocols that the ENS uses.
By encapsulating fuzzing inputs according to the identified protocol tree, \toolname allows the fuzzer to test the implementation of different layers of the ENS, as well as the application logic of the firmware.

To demonstrate the versatility and applicability of our approach, we integrate \toolname with three popular rehosting-based firmware fuzzers: \fuzzware~\cite{scharnowski2022fuzzware}, \hoedur~\cite{scharnowski2023hoedur}, and \semu~\cite{zhou2022semu}.
We then evaluate \toolname's performance on a firmware sample set that covers a wide variety of network protocols.
Our results show that \toolname consistently improves code coverage for the tested firmware samples. More specifically, our approach increases the average basic block coverage by 40.7\% when used together with \fuzzware, by 39.2\% with \hoedur, and by 8.5\% with \semu.
Moreover, we evaluated \toolname against \emnettest, the only other technique specifically targeting ENS testing. Using \emnettest's dataset of 12 real-world vulnerabilities, we show that \toolname not only successfully rediscovered all of them, but \toolname also identified three additional bugs missed by \emnettest. Combined with two further novel bugs---one of them in the \semu dataset---\toolname managed to discover five new bugs in network-exposed code throughout our evaluation.
We are in the process of disclosing all of our findings to the affected vendors in a coordinated way and one of them has been fixed.

\smallskip
\noindent
\textbf{Contributions.}
We provide the following three key contributions:
\begin{compactitem}
\item We present a new approach for dynamically detecting the use of network protocols in embedded firmware. Based on the identified network interactions, we propose a method to encapsulate fuzzing data in valid network packets based on a \emph{virtual network} to effectively test different layers of network interactions.
\item To demonstrate the general applicability of our method across different fuzzing frameworks, we integrate our approach into three frameworks: \fuzzware, \hoedur, and \semu. 
\item In a comprehensive ablation study, we show that \toolname significantly improves code coverage across all application scenarios and outperforms existing techniques on vulnerability benchmarks by finding five previously unknown bugs. 
\end{compactitem}

To foster research on this topic, we publish the source code, our sample data set, and other research artifacts at \url{https://github.com/MPI-SysSec/pemu}.

\section{Background}
We first provide a brief overview of firmware and the specific challenges involved in analyzing the security of their network stacks.

\subsection{Firmware in Embedded Systems}
Embedded systems tightly integrate hardware and software to perform specific tasks within devices ranging from IoT gadgets to industrial systems~\cite{papp_embedded_2015, heath_embedded_2002}. Unlike general-purpose computers, they typically contain a microcontroller unit (MCU). This self-contained system typically includes a processor, memory, and various on-chip peripherals such as Serial Peripheral Interface (SPI), Universal Asynchronous Receiver-Transmitter (UART), or Direct Memory Access (DMA). These peripherals allow connectivity to off-chip components like Ethernet controllers, Bluetooth modules, wireless modems, or LED screens. The wide range of customized processors and peripherals, both on-chip and off-chip, adds to the complexity and diversity of these systems.

Firmware, the software controlling an embedded system, is usually built on either a minimal embedded operating system (EOS) or directly on the hardware (referred to as ``bare metal''), resulting in fewer abstraction layers than in general-purpose software. 
Furthermore, embedded applications are strongly integrated with either the EOS or the system's underlying hardware.
Firmware components, such as network stacks and libraries, are specifically tailored to the hardware and closely integrated, which makes standard security analyses challenging and calls for customized techniques.

\subsection{Linux-based Fuzzing}
\emph{Fuzzing} is a well-established security testing technique that generates semi-random inputs to identify unexpected or crash-inducing behavior in a system~\cite{manes2019art,AFLplusplus-Woot20}. However, fuzzing embedded systems on real hardware is often impractical due to poor scalability and limited observability~\cite{muench2018wycinwyc}. For Linux-based, type 1~\cite{muench2018wycinwyc} devices like routers or IP cameras, previous research has focused on
fuzzing user-space applications. This is either done by emulating the target application in QEMU user-mode~\cite{tay2023greenhouse} or on top of a customized kernel in QEMU system-mode~\cite{chen2016towards, kim2020firmae}.
In both cases, the primary focus of these approaches is setting up the target's Linux-related environment, such as the file system or network interfaces. To this end, they use well-known kernel interfaces, like system calls, to collect data such as required files and IP addresses.

\subsection{Rehosting-based Fuzzing}
In contrast, applications running on tightly integrated type 2 and type 3 EOSs~\cite{muench2018wycinwyc} cannot be easily emulated by these Linux-based approaches. This is because these EOSs do not provide the same high-level abstractions that Linux provides to mask the underlying hardware complexity. Instead, \emph{firmware rehosting} addresses the challenge of embedded fuzzing by emulating bare-metal or EOS-based firmware on commodity hardware, enabling the use of dynamic analysis techniques that are otherwise unfeasible on embedded devices. By decoupling the firmware from specific hardware dependencies, rehosting provides a more flexible and scalable platform for security analysis.
Recent approaches have further reduced the need for exact peripheral emulation by using generic abstraction models~\cite{clements2020halucinator}, automatic peripheral interaction modeling~\cite{feng2020p2im, scharnowski2022fuzzware}, and customized input formats~\cite{scharnowski2023hoedur, chesser2024multifuzz}. These techniques support effective fuzzing by providing raw input to the firmware through modeled peripherals.

\subsection{Fuzzing Embedded Network Stacks}
\begin{figure}[tb]
    \centering
    \includegraphics[width=.93\columnwidth]{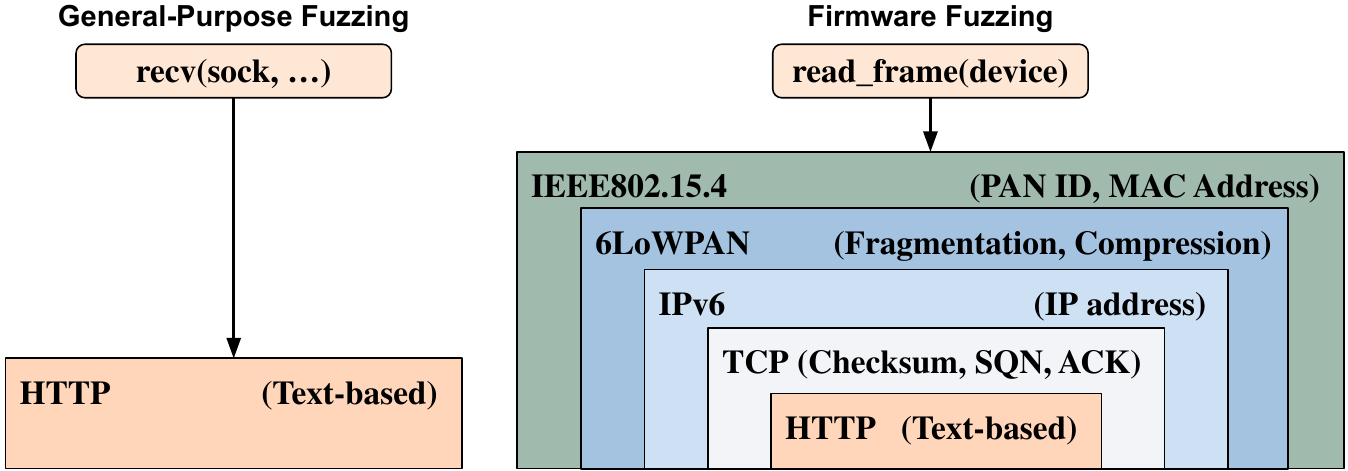}
    \caption{Comparison of the input formats required for testing an HTTP server implementation between general-purpose fuzzing and firmware fuzzing: given the lack of abstractions provided by an OS, a firmware fuzzer needs to deal with all individual network layers.}
    \label{fig:background-HTTP-fuzzing}
\end{figure}

While rehosting-based fuzzing enables broad firmware testing, it still falls short when dealing with more complex components, especially network stacks and the applications running on top of them.
Network stacks consist of multiple interdependent layers, and data is passed to higher layers using complex formats and specific sequences.
In general-purpose systems, user-space fuzzing of network services benefits from OS-level abstractions, e.g., a fuzzer can inject test cases directly into a program via system calls like \texttt{recv}, without needing to construct the underlying network protocol layers.  In contrast, embedded firmware fuzzing lacks such abstractions.
As illustrated in Figure~\ref{fig:background-HTTP-fuzzing}, this lack of abstraction makes fuzzing far more complex. Instead of directly testing the application layer, the fuzzer must provide input that satisfies \emph{all} underlying network layers (i.e., from IEEE 802.15.4 to 6LoWPAN and IPv6 up to TCP and the actual HTTP payload).
This means the fuzzer must generate inputs that satisfy the expectations of every layer (from the data link layer up to the application) without any OS assistance. 
In contrast, Linux-based fuzzing uses readily available network interfaces provided by the Linux kernel to send application data directly to the target process~\cite{kim2020firmae, green2022graphfuzz, chen2016towards}.

As current state-of-the-art fuzzers and rehosting frameworks offer no notion of message encapsulation or internal state, they are ill-equipped to test such complex components. A fuzzer with a focus on bitflips is unlikely to craft input that resembles a nested packet valid on different layers, each with its unique constraints. At the same time, existing fuzzers have no notion of statefulness between messages, nor do they explicitly account for values that must be maintained across individual messages.

\section{Design}%
\label{chapter:design}

While fuzzing network applications is well studied for general-purpose systems, embedded systems pose unique challenges that remain largely unaddressed.
State-of-the-art firmware fuzzers typically provide only raw data to the network stack, which fails to match the complex syntax and semantics of network protocols.
As a result, the network stack often discards packets generated during fuzzing before they can reach deeper code sections, including custom application logic. This limits the fuzzer’s effectiveness and leads to repetitive, ineffective attempts to bypass lower-layer constraints such as checksums, static addresses, or incrementing values.
Existing firmware fuzzers do not effectively address this limitation. 
Using captured network traffic as seeds, as \aflnet does, has several drawbacks. Rehosting is supposed to work without physical hardware, so relying on real traffic undermines this benefit. Additionally, mutating such seeds regularly breaks checksums or protocol structures. Fixing this either requires source code access (which rehosting avoids) or checksum recalculation after every mutation, making it rather expensive.

To overcome these limitations, we present the design of \toolname, a virtual network that delivers realistic, syntactically valid packets to embedded firmware under test. Through a transparent and self-configuring encapsulation mechanism, \toolname allows the rehosting platform to request and deliver well-formed network packets just in time. \toolname also analyzes firmware-transmitted packets, from which it can extract important network values, such as addresses and sequence numbers.
By enforcing both syntactic and semantic correctness at each network layer, \toolname enables fuzzing at the application layer, bypassing lower-layer constraints like checksum and header validation. 
Based on this foundation, we introduce two automated analysis methods that identify protocols and protocol-specific values in the firmware \emph{without} requiring prior knowledge or manual intervention.

\subsection{High-Level Overview}
\begin{figure}[tb]
    \centering
    \includegraphics[width=.9\columnwidth]{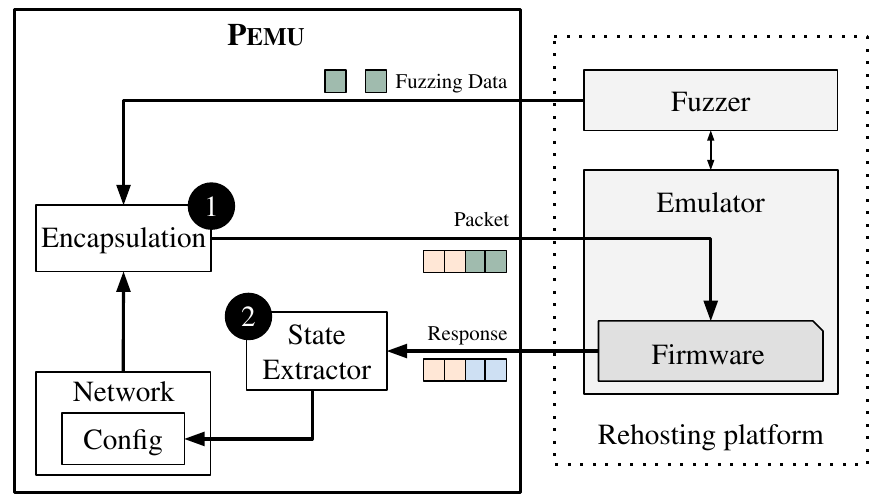}
    \caption{High-Level Overview of \toolname's architecture}
    \label{fig:overview}
\end{figure}

From the firmware's perspective, \toolname dynamically creates a \vn that facilitates valid network interactions between the firmware and a fuzzing-based rehosting platform.
To achieve this, \toolname consists of multiple components (see Figure~\ref{fig:overview}) that interact with each other.
When the rehosting platform detects that the firmware is in a state where it can receive a packet, the \packer generates such a packet by using raw fuzzing input from the rehosting platform's fuzzing module. It then encapsulates this input into a valid network packet based on the current (initially empty) network configuration (\circleone).
Guided by the fuzzer, the \packer adjusts how deeply it encapsulates the input, allowing targeted testing of different layers in the network stack.
Once the firmware transmits low-level frames, \toolname parses these frames to extract information about the current protocol state, \eg source and destination addresses (\circletwo).
This state is later used by the \packer to encapsulate network packets correctly.
The network is initialized with an empty configuration to facilitate automated end-to-end testing. Through iterative analysis steps, \toolname dynamically derives the protocol stack used by the firmware by using protocol-specific probes and code coverage as feedback. This probing process is repeated throughout the fuzzing campaign to continuously refine the configuration as new behavior of the firmware's network stack is discovered (see Section~\ref{sec:design:probing}).

\subsection{Protocol-Aware Firmware Rehosting}
We now discuss the main components of \toolname in detail and explain how the encapsulation and extraction steps work.

\subsubsection{Packet Encapsulation}
Given the current network configuration, the fuzzer can utilize the \packer to encapsulate raw input into realistic network packets. 
Varying the depth of the encapsulation in a fuzzer-driven manner allows \toolname to equally test every layer of the network stack, unlike most existing general-purpose network application fuzzers.
By generating semantically correct packets, the \packer acts as a funnel that enables the fuzzer to target specific firmware components within the network stack dynamically.

The underlying insight is that all network packets follow a layered architecture: each protocol within a packet consists of a header with metadata and a body. Only the body of the highest layer contains the actual application data, while the body of each lower layer encapsulates the layer immediately above it.
Header fields may contain static values (e.g., \texttt{version = 4} in IPv4), references to another layer (e.g., \texttt{EtherType = 8} in Ethernet when the next layer is IPv4), or dynamically computed fields such as checksums. We supply the \packer with a set of protocol grammars to translate these constraints into raw network packets. 
We use the official specifications to manually derive these grammars (which is a one-time effort). Our grammar can depict complex constraints like fragmentation, compression, checksums, or static values. These protocol grammars allow the \packer to create valid packets that contain fuzzing input as the body of a given encapsulating protocol. To enable broad testing of each protocol layer, we designed the grammar to use fuzzing input for every header field, which is not bound by strict semantic requirements (\eg addresses and checksums).

The \packer often has to send multiple low-level frames, for instance, due to fragmentation or to perform a handshake before any data reaches
the firmware's application layer. Hence, in addition to correctly assembling individual packets, the \packer must maintain state information across a sequence of packets, such as sequence numbers, TCP SYN/SYN-ACK pairs, or fragmentation metadata.
This highlights one of the core insights of \toolname: While the firmware might successfully parse a single network packet consisting of only random data by chance, the likelihood of two subsequent randomly generated packets being accepted as a valid application-level input is very low. As a result, current firmware fuzzers are often unable to progress up to the application layer of firmware, as they lack an understanding of the underlying protocol semantics.
Furthermore, based on fuzzing input, the \packer can 
mutate individual header fields. Introducing mutations into the packet allows us to test the boundaries of the network stack. This approach is motivated by the findings of Amusuo \etal, who found that 95\% of vulnerabilities in embedded network stacks depend on only one or two incorrect header fields~\cite{amusuo2023systematically}.

\subsubsection{State Extraction}
To improve the encapsulation process, we analyze \emph{packets that are sent by} the firmware to derive values related to the network state. While existing rehosting platforms largely ignore the firmware output, \toolname actively uses these output packets to inform the input encapsulation performed by the \packer.
There are two reasons for firmware network interactions that we aim to capture protocol state from, \emph{unsolicited} and \emph{solicited} packets. \emph{Unsolicited packets} are packets the firmware network stack sends without any external trigger. The sending occurs either during the firmware's initial setup process or periodically. In the case of a TCP/IP stack, examples are ARP packets to identify the gateway's MAC address, DHCP packets to request an IP address, and ICMP pings to verify connectivity. The firmware includes header metadata of all involved protocols in all cases, e.g., an ARP packet contains the MAC address as well as the sender's IP address.
In contrast, \emph{solicited packets} are prompted in response to a packet the firmware received. Examples include the firmware sending a TCP SYN-ACK segment with a sequence number in response to a TCP SYN segment or requesting an IP address in response to a DHCP offer packet. 

Whenever the firmware sends a packet, the \parser module extracts the included protocol state values so they can be added to the configuration of the \vn accordingly. These values are then used when encapsulation fuzzing input, as described in the previous section.
This approach eliminates the guesswork that a fuzzer would otherwise face when constructing sequences of packets with valid metadata across multiple protocol layers.

\subsection{Platform Independency}
\toolname is designed to be independent of the underlying rehosting platform, requiring only a minimal adapter for integration. This adapter enables communication between the platform and \toolname, while the platform itself remains responsible for deciding when to request packets and where to deliver them. 
This intentional choice allows \toolname to remain agnostic to the specific emulation context and the methods used for input handling. Fundamentally, the platform must provide \toolname only access to fuzzer-generated input (for packet encapsulation). It can then leverage \toolname's straightforward API of two commands: \texttt{get\_packet} and \texttt{send\_packet}. 
To showcase our design's generic applicability, we implement it on top of three different rehosting platforms (see Section~\ref{sec:implementation}).
 
\subsection{\toolname-based Network Traffic Analysis}
Based on \toolname's encapsulation and state extraction, we design two analysis methods that enable the underlying rehosting platform to perform an end-to-end analysis of a target \emph{without} requiring any manual effort or previous knowledge to extract a network configuration. Furthermore, the gained information can be used to broaden \toolname's knowledge of the firmware's network stack.
These passes are designed to be run while the fuzzer is paused to ensure a coherent configuration.

\subsubsection{Packet Sequence Extraction}
\label{sec:design:probing}

For large-scale analyses, avoiding manually reverse-engineering individual samples to extract information on the type of network packets the firmware accepts is impractical
desirable. Manual analysis is not only time-consuming and labor-intensive but also error-prone and difficult to scale across large and diverse firmware samples. Hence, we propose an end-to-end approach that can extract this information without any manual involvement is desirable.
We introduce a technique called \emph{coverage-based probing} to achieve this goal. The underlying idea is the following: We expect the firmware implementation of a given network layer to react distinctly to well-formed input, \ie a well-formed input triggers unique coverage when compared to a malformed input. By generating a set of well-formed inputs for each network layer candidate, we can observe which packets elicit a distinct reaction from the firmware. The packets that trigger such distinct reactions likely represent the network protocols used by the firmware under test.

There are several ways to measure these reactions based on firmware coverage. The naive approach of only selecting the packet types that cover the highest number of basic blocks has a significant drawback: The error-handling routine, which is likely triggered by a malformed input, produces coverage. We even found that the number of basic blocks that are part of error handling can be  higher than the amount of ``regular'' coverage induced by a correct packet.
To address this issue, we use the metric of \emph{uniquely covered basic blocks} instead of the total number of basic blocks covered.
This approach is based on the insight that well-formed packets of the expected type will progress further into the network stack, while unexpected packets will eventually trigger some common error-handling routine. 
Hence, valid packet types will likely cover the highest number of unique basic blocks.

A challenging aspect of automatically identifying the network configuration is that protocols in the network stack rely on other protocols, even protocols on the same topological layer. More specifically, often,  before the firmware can receive a packet containing a given protocol, separate packets of an entirely different protocol may have to be exchanged first. An example of this is ARP and IPv4, where ARP is used to identify the MAC address of a gateway so that the firmware can further communicate with it via IPv4. Similarly, a firmware that implements a web server via TCP may need to dynamically configure its IPv4 address via DHCP, which relies on UDP, before it can accept TCP segments.

\colorlet{rubcolor}{black}
\definecolor{blizzardblue}{rgb}{0.67, 0.9, 0.93}
\definecolor{bluegray}{HTML}{a4c3e0}
\definecolor{special_o}{HTML}{ffd6ba}
\colorlet{tikznewcolor}{special_o}
\colorlet{tikzhlcolor}{bluegray}

\begin{figure}[tb]
    \centering
\noindent\begin{subfigure}{0.5\columnwidth}
\scriptsize\centering%
    \resizebox{0.9\textwidth}{!}{%
    \fbox{%
	\begin{tikzpicture}[%
			defaultnode/.style={text height=1em, text centered, minimum width=3em},
            hlnode/.style={rectangle, thick, draw=tikzhlcolor!90, fill=tikzhlcolor!10,  text height=1em, text centered, minimum width=3em, },
            newnode/.style={rectangle, thick, draw=tikznewcolor!90, fill=tikznewcolor!10,  text height=1em, text centered, minimum width=3em, },
            every path/.style={-{Latex[length=0.9mm, width=0.6mm]}}
		]

	\node[newnode] (top) {\vphantom{Hy}\hphantom{F}Frame\hphantom{F}};
	\node[defaultnode] (secondtwo)   [below = 1em of top]         {\vphantom{Hy}\dots};
	\node[defaultnode] (secondone)   [left = 1em of secondtwo]    {\vphantom{Hy}Ethernet};
	\node[defaultnode] (secondthree) [right = 1em of secondtwo]   {\vphantom{Hy}BLE};

	\node[defaultnode] (thirdtwo)   [below = 1em of secondone]  {\vphantom{Hy}IPv4};
	\node[defaultnode] (thirdone)   [left = 1em of thirdtwo]   {\vphantom{Hy}ARP};
	\node[defaultnode] (thirdthree) [right = 1em of thirdtwo]  {\vphantom{Hy}\dots};
	\node[defaultnode] (thirdfour) [below = 1em of secondthree]  {\vphantom{Hy}\dots};

	\node[draw=none, fill=none, text height= 1em] (fourthanchor) [below = 1em of thirdtwo]       {\vphantom{Hy}};
	\node[defaultnode] (fourthone)                               [left = 0.1em of fourthanchor]  {\vphantom{Hy}UDP};
	\node[defaultnode] (fourthtwo)                               [right = 2.5em of fourthanchor] {\vphantom{Hy}TCP};

	\node[draw=none, fill=none, text height= 1em] (fifthanchor) [below = 1em of fourthone]     {\vphantom{Hy}};
	\node[defaultnode] (fifthone)                               [left = 0em of fifthanchor]  {\vphantom{Hy}DHCP};
	\node[defaultnode] (fifthtwo)                               [right = 0em of fifthanchor] {\vphantom{Hy}SNMP};
	\node[defaultnode] (fifththree)                             [below = 1em of fourthtwo] {\vphantom{Hy}MQTT};

	\draw (top.210) -- node {} ++(0,-0.5em) -| (secondone.north);
	\draw (top.270) --                         (secondtwo.north);
	\draw (top.330) -- node {} ++(0,-0.5em)   -| (secondthree.north);
	
	\draw (secondone.210) -- node {} ++(0,-0.5em) -| (thirdone.north);
	\draw (secondone.270) --                         (thirdtwo.north);
	\draw (secondone.330) -- node {} ++(0,-0.5em) -| (thirdthree.north);
	\draw (secondthree.270) --                         (thirdfour.north);

	\draw (thirdtwo.235) -- node {} ++(0,-0.5em)   -| (fourthone.north);
	\draw (thirdtwo.305) -- node {} ++(0,-0.5em)   -| (fourthtwo.north);

	\draw (fourthone.235) -- node {} ++(0,-0.5em)   -|  (fifthone.north);
	\draw (fourthone.305) -- node {} ++(0,-0.5em)   -|  (fifthtwo.north);
	\draw (fourthtwo.270) --                          (fifththree.north);
		
\end{tikzpicture} %
}%
}%
\caption{Initial network configuration}
\end{subfigure}%
\hfill%
\noindent\begin{subfigure}{0.5\columnwidth}%
\scriptsize\centering%
    \resizebox{0.9\textwidth}{!}{%
    \fbox{%
	\begin{tikzpicture}[%
			defaultnode/.style={text height=1em, text centered, minimum width=3em},
            hlnode/.style={rectangle, thick, draw=tikzhlcolor!90, fill=tikzhlcolor!10,  text height=1em, text centered, minimum width=3em, },
            newnode/.style={rectangle, thick, draw=tikznewcolor!90, fill=tikznewcolor!10,  text height=1em, text centered, minimum width=3em, },
            every path/.style={-{Latex[length=0.9mm, width=0.6mm]}}
		]

	\node[hlnode] (top) {\vphantom{Hy}\hphantom{F}Frame\hphantom{F}};
	\node[defaultnode] (secondtwo)   [below = 1em of top]         {\vphantom{Hy}\dots};
	\node[newnode] (secondone)   [left = 1em of secondtwo]    {\vphantom{Hy}Ethernet};
	\node[defaultnode] (secondthree) [right = 1em of secondtwo]   {\vphantom{Hy}BLE};

	\node[defaultnode] (thirdtwo)   [below = 1em of secondone]  {\vphantom{Hy}IPv4};
	\node[newnode] (thirdone)   [left = 1em of thirdtwo]   {\vphantom{Hy}ARP};
	\node[defaultnode] (thirdthree) [right = 1em of thirdtwo]  {\vphantom{Hy}\dots};
	\node[defaultnode] (thirdfour) [below = 1em of secondthree]  {\vphantom{Hy}\dots};

	\node[draw=none, fill=none, text height= 1em] (fourthanchor) [below = 1em of thirdtwo]       {\vphantom{Hy}};
	\node[defaultnode] (fourthone)                               [left = 0.1em of fourthanchor]  {\vphantom{Hy}UDP};
	\node[defaultnode] (fourthtwo)                               [right = 2.5em of fourthanchor] {\vphantom{Hy}TCP};

	\node[draw=none, fill=none, text height= 1em] (fifthanchor) [below = 1em of fourthone]     {\vphantom{Hy}};
	\node[defaultnode] (fifthone)                               [left = 0em of fifthanchor]  {\vphantom{Hy}DHCP};
	\node[defaultnode] (fifthtwo)                               [right = 0em of fifthanchor] {\vphantom{Hy}SNMP};
	\node[defaultnode] (fifththree)                             [below = 1em of fourthtwo] {\vphantom{Hy}MQTT};

	\draw (top.210) -- node {} ++(0,-0.5em) -| (secondone.north);
	\draw (top.270) --                         (secondtwo.north);
	\draw (top.330) -- node {} ++(0,-0.5em)   -| (secondthree.north);
	
	\draw (secondone.210) -- node {} ++(0,-0.5em) -| (thirdone.north);
	\draw (secondone.270) --                         (thirdtwo.north);
	\draw (secondone.330) -- node {} ++(0,-0.5em) -| (thirdthree.north);
	\draw (secondthree.270) --                         (thirdfour.north);

	\draw (thirdtwo.235) -- node {} ++(0,-0.5em)   -| (fourthone.north);
	\draw (thirdtwo.305) -- node {} ++(0,-0.5em)   -| (fourthtwo.north);

	\draw (fourthone.235) -- node {} ++(0,-0.5em)   -|  (fifthone.north);
	\draw (fourthone.305) -- node {} ++(0,-0.5em)   -|  (fifthtwo.north);
	\draw (fourthtwo.270) --                          (fifththree.north);
\end{tikzpicture}    %
}%
}%
\caption{Iteration 1: ARP detected}%
\label{design:config-iteration:b}
\end{subfigure}%

\noindent\begin{subfigure}{0.5\columnwidth}
\scriptsize\centering%
    \resizebox{0.9\textwidth}{!}{%
    \fbox{%
	\begin{tikzpicture}[%
			defaultnode/.style={text height=1em, text centered, minimum width=3em},
            hlnode/.style={rectangle, thick, draw=tikzhlcolor!90, fill=tikzhlcolor!10,  text height=1em, text centered, minimum width=3em, },
            newnode/.style={rectangle, thick, draw=tikznewcolor!90, fill=tikznewcolor!10,  text height=1em, text centered, minimum width=3em, },
            every path/.style={-{Latex[length=0.9mm, width=0.6mm]}}
		]

	\node[hlnode] (top) {\vphantom{Hy}\hphantom{F}Frame\hphantom{F}};
	\node[defaultnode] (secondtwo)   [below = 1em of top]         {\vphantom{Hy}\dots};
	\node[hlnode] (secondone)   [left = 1em of secondtwo]    {\vphantom{Hy}Ethernet};
	\node[defaultnode] (secondthree) [right = 1em of secondtwo]   {\vphantom{Hy}BLE};

	\node[newnode] (thirdtwo)   [below = 1em of secondone]  {\vphantom{Hy}IPv4};
	\node[hlnode] (thirdone)   [left = 1em of thirdtwo]   {\vphantom{Hy}ARP};
	\node[defaultnode] (thirdthree) [right = 1em of thirdtwo]  {\vphantom{Hy}\dots};
	\node[defaultnode] (thirdfour) [below = 1em of secondthree]  {\vphantom{Hy}\dots};

	\node[draw=none, fill=none, text height= 1em] (fourthanchor) [below = 1em of thirdtwo]       {\vphantom{Hy}};
	\node[newnode] (fourthone)                               [left = 0.1em of fourthanchor]  {\vphantom{Hy}UDP};
	\node[defaultnode] (fourthtwo)                               [right = 2.5em of fourthanchor] {\vphantom{Hy}TCP};

	\node[draw=none, fill=none, text height= 1em] (fifthanchor) [below = 1em of fourthone]     {\vphantom{Hy}};
	\node[newnode] (fifthone)                               [left = 0em of fifthanchor]  {\vphantom{Hy}DHCP};
	\node[defaultnode] (fifthtwo)                               [right = 0em of fifthanchor] {\vphantom{Hy}SNMP};
	\node[defaultnode] (fifththree)                             [below = 1em of fourthtwo] {\vphantom{Hy}MQTT};

	\draw (top.210) -- node {} ++(0,-0.5em) -| (secondone.north);
	\draw (top.270) --                         (secondtwo.north);
	\draw (top.330) -- node {} ++(0,-0.5em)   -| (secondthree.north);
	
	\draw (secondone.210) -- node {} ++(0,-0.5em) -| (thirdone.north);
	\draw (secondone.270) --                         (thirdtwo.north);
	\draw (secondone.330) -- node {} ++(0,-0.5em) -| (thirdthree.north);
	\draw (secondthree.270) --                         (thirdfour.north);

	\draw (thirdtwo.235) -- node {} ++(0,-0.5em)   -| (fourthone.north);
	\draw (thirdtwo.305) -- node {} ++(0,-0.5em)   -| (fourthtwo.north);

	\draw (fourthone.235) -- node {} ++(0,-0.5em)   -|  (fifthone.north);
	\draw (fourthone.305) -- node {} ++(0,-0.5em)   -|  (fifthtwo.north);
	\draw (fourthtwo.270) --                          (fifththree.north);
		
\end{tikzpicture} %
}%
}%
\caption{Iteration 2: DHCP detected}
\end{subfigure}%
\hfill%
\noindent\begin{subfigure}{0.5\columnwidth}%
\scriptsize\centering%
    \resizebox{0.9\textwidth}{!}{%
    \fbox{%
	\begin{tikzpicture}[%
			defaultnode/.style={text height=1em, text centered, minimum width=3em},
            hlnode/.style={rectangle, thick, draw=tikzhlcolor!90, fill=tikzhlcolor!10,  text height=1em, text centered, minimum width=3em, },
            newnode/.style={rectangle, thick, draw=tikznewcolor!90, fill=tikznewcolor!10,  text height=1em, text centered, minimum width=3em, },
            every path/.style={-{Latex[length=0.9mm, width=0.6mm]}}
		]

	\node[hlnode] (top) {\vphantom{Hy}\hphantom{F}Frame\hphantom{F}};
	\node[defaultnode] (secondtwo)   [below = 1em of top]         {\vphantom{Hy}\dots};
	\node[hlnode] (secondone)   [left = 1em of secondtwo]    {\vphantom{Hy}Ethernet};
	\node[defaultnode] (secondthree) [right = 1em of secondtwo]   {\vphantom{Hy}BLE};

	\node[hlnode] (thirdtwo)   [below = 1em of secondone]  {\vphantom{Hy}IPv4};
	\node[hlnode] (thirdone)   [left = 1em of thirdtwo]   {\vphantom{Hy}ARP};
	\node[defaultnode] (thirdthree) [right = 1em of thirdtwo]  {\vphantom{Hy}\dots};
	\node[defaultnode] (thirdfour) [below = 1em of secondthree]  {\vphantom{Hy}\dots};

	\node[draw=none, fill=none, text height= 1em] (fourthanchor) [below = 1em of thirdtwo]       {\vphantom{Hy}};
	\node[hlnode] (fourthone)                               [left = 0.1em of fourthanchor]  {\vphantom{Hy}UDP};
	\node[newnode] (fourthtwo)                               [right = 2.5em of fourthanchor] {\vphantom{Hy}TCP};

	\node[draw=none, fill=none, text height= 1em] (fifthanchor) [below = 1em of fourthone]     {\vphantom{Hy}};
	\node[hlnode] (fifthone)                               [left = 0em of fifthanchor]  {\vphantom{Hy}DHCP};
	\node[defaultnode] (fifthtwo)                               [right = 0em of fifthanchor] {\vphantom{Hy}SNMP};
	\node[newnode] (fifththree)                             [below = 1em of fourthtwo] {\vphantom{Hy}MQTT};

	\draw (top.210) -- node {} ++(0,-0.5em) -| (secondone.north);
	\draw (top.270) --                         (secondtwo.north);
	\draw (top.330) -- node {} ++(0,-0.5em)   -| (secondthree.north);
	
	\draw (secondone.210) -- node {} ++(0,-0.5em) -| (thirdone.north);
	\draw (secondone.270) --                         (thirdtwo.north);
	\draw (secondone.330) -- node {} ++(0,-0.5em) -| (thirdthree.north);
	\draw (secondthree.270) --                         (thirdfour.north);

	\draw (thirdtwo.235) -- node {} ++(0,-0.5em)   -| (fourthone.north);
	\draw (thirdtwo.305) -- node {} ++(0,-0.5em)   -| (fourthtwo.north);

	\draw (fourthone.235) -- node {} ++(0,-0.5em)   -|  (fifthone.north);
	\draw (fourthone.305) -- node {} ++(0,-0.5em)   -|  (fifthtwo.north);
	\draw (fourthtwo.270) --                          (fifththree.north);
\end{tikzpicture}    %
}%
}%
\caption{Iteration 3: MQTT detected}
\end{subfigure}%
    \caption{Exemplary space of available network protocols and their nesting, from the point of view of \toolname (\begin{tikzpicture}\protect\node[rectangle, fill=tikznewcolor!10, draw=tikznewcolor, line width=1pt, text width=0.3em] (n) {};\end{tikzpicture} newly detected, \begin{tikzpicture}\protect\node[rectangle, fill=tikzhlcolor!10, draw=tikzhlcolor,line width=1pt, text width=0.3em] (n) {};\end{tikzpicture} previously detected). \toolname successively detects more and more protocol layers and adds them to the detected network configuration at runtime.}%
    \label{fig:design:config-iteration}
\end{figure}

Building on these insights, we devise an iterative approach for detecting the entire stack of network protocols used by the firmware. For a given state of fuzzing progress, \toolname detects the next network layer currently reachable. This newly identified layer is then added to the existing network configuration, \ie the set of protocol layers currently assumed to be used by the firmware. This configuration is usually empty at the beginning of the fuzzing campaign. By incrementally updating the network configuration, our approach allows the fuzzing to progress deeper into the network stack. 

Figure~\ref{fig:design:config-iteration} shows how the configuration is updated over three iterations for an example firmware that exposes an MQTT server. Initially, no protocol layers have been identified (Figure~\ref{fig:design:config-iteration}a). In this initial configuration, fuzzing input is passed into raw frames. This state is identical to how existing rehosting platforms pass fuzzing input without \toolname's \vn. The example firmware starts resolving the MAC address of its gateway. To this end, the ENS implementation sends ARP requests, expecting an ARP response. In this state, a probe packet containing a valid ARP response triggers the network stack to store the MAC address of its gateway successfully. This event leads to unique code coverage compared to other invalid probe requests (which are discarded by the corresponding layers). Consequently, \toolname adds the Ethernet and ARP protocols to its network configuration, resulting in Iteration~1 (see Figure~\ref{fig:design:config-iteration}b). Similarly, once the example firmware requests an IPv4 address via DHCP, a probe request that contains a valid DHCP packet will generate unique coverage further up the firmware's network stack. This progression leads to Iteration~2 (Figure~\ref{fig:design:config-iteration}c). Finally, after the IPv4 address is assigned, the firmware will start listening for incoming TCP connections to serve MQTT requests. This behavior can again be detected by probe requests, leading to Iteration~3 of the configuration (Figure~\ref{fig:design:config-iteration}d).

An appealing property of this iterative, incremental approach is its \emph{robustness}: Even in the unlikely case that a probe request mistakenly adds an unused protocol layer into the detected network configuration, this protocol is given to the fuzzer as an \emph{option} for encapsulation. As the fuzzer decides up to which layer to encapsulate, it will not be forced to send a wrong type of network packet continuously. In fact, coverage feedback will lead the fuzzer to deprioritize wrong choices.

\subsubsection{Parsing-Based Analysis}
The concept of parsing-based analysis refers to the comprehensive analysis of the results of the state extraction provided by \toolname. First, all extracted values are collected and analyzed. When multiple options exist for a header field, the most likely one is chosen and integrated into \toolname's configuration for the next fuzzing run. This iterative and self-correcting approach significantly improves the quality of generated network packets, while reducing the required fuzzing input. 

\section{Implementation}%
\label{sec:implementation}

To evaluate our approach, we implemented a prototype of our design. 
Additionally, we integrated our approach with three state-of-the-art rehosting platforms: \emph{\fuzzware}~\cite{scharnowski2022fuzzware}, \emph{\hoedur}~\cite{scharnowski2023hoedur}, and \emph{\semu}~\cite{zhou2022semu}. We release \toolname as well as our patches to these platforms at \url{https://github.com/MPI-SysSec/pemu}.

\subsection{Implementation Aspects of \toolname}
\toolname's \vn exposes two core interfaces: one for receiving packets from the network and one for sending packets to the network. We implemented \toolname in Python, which is supported by most rehosting platforms. The implementation consists of about 3,600 lines of Python code.

\paragraph{\bfseries \Packer}
The \vn provides a dedicated module, the \packer, which monitors the current network state and assembles packets as needed.
The \packer is initialized with a predefined list of network packets (\ie the network configuration). Upon each request from the rehosting platform, it processes the list and assembles the packet at the current index before advancing to the next.
Protocol semantics and syntax are defined in YAML configuration files and are derived from official specifications. Header fields are classified into the following categories: 
\begin{itemize}
    \item Static values that are specified directly in the configuration. 
    \item Length values and the scope they refer to. 
    \item Fields that reference higher layers
    \item Fields that link handler functions for complex calculations, like checksums.
    \item State-related values that require updates, such as flags and sequence numbers.
    \item Remaining fields that can be populated with fuzzing data.
\end{itemize}

Like an actual network stack, the \packer assembles packets layer by layer, starting from the highest to the lowest protocol layer. 
Additionally, the \vn supports fuzzer-guided fault injection, allowing it to introduce mutations into packets to test edge cases and error-handling logic. These malformed packets can help uncover vulnerabilities in lower-layer parsers that might be missed with well-formed inputs alone.
Alongside our tool, we publish YAML configuration files for 38 protocols.

\paragraph{\bfseries  State Extraction}
The State Extraction module parses outgoing packets according to their protocol specifications. The module identifies and stores unknown protocol values (\eg IP addresses, IDs, and nonces) for later analysis. Complex header fields, such as the Frame Control Field (FCF) in IEEE 802.15.4 or cases involving fragmentation and compression, can be managed with custom handler functions provided by the user. Adding handler functions is a one-time effort when adding a new protocol. 

\subsection{\toolname-based Analysis}
We also implemented two complementary analysis techniques that enable end-to-end automation without manual intervention. These analyses are performed after fuzzing is paused, and their results are then integrated into the network configuration upon continuation. This is done to prevent inconsistent configurations.

\paragraph{\bfseries Analyzing Extracted State}
During fuzzing, the \vn's \parser module inspects and extracts relevant information from packets transmitted by the firmware. This data is accumulated and analyzed to refine and expand the network configuration with new information on connection-specific values and new network packet types, enhancing future fuzzing effectiveness.

\paragraph{\bfseries Probing}
Relying solely on state extraction to explore the network state has limitations, particularly when there is no prior knowledge of the protocol suite or when the firmware, based on its role (e.g., client or server), remains idle without an external prompt.  We use an active probing approach to address this issue, as detailed in  Section~\ref{sec:design:probing}. This method leverages heuristics to select probes, or packet types, that are most likely to advance exploration of the firmware’s network stack.
To this end, the \packer supports a probe mode, where it generates packets with randomized yet syntactically valid configurations based on the current firmware state. All fuzzing bytes in the probe packets are set to zero to ensure consistency across probing runs and avoid non-deterministic behavior from random inputs. Only non-fuzzed fields—such as those derived from static values, handlers, checksums, lengths, and pointers—remain unaltered, reducing the risk of random bytes misleadingly mimicking unintended protocol fields. This ensures, for example, that a randomized Ethernet MAC address is not erroneously interpreted as an IEEE 802.15.4 frame.

Subsequently, the probing component places the \packer in the dedicated probe mode, extracts the gathered coverage, and applies the heuristics to the collected basic block sets.
After running the probes, coverage data is collected, and heuristics are applied to assess the firmware’s basic block responses. The probing module includes a communication interface adaptable to different rehosting platforms, ensuring precise extraction and interpretation of coverage data.

\subsection{Integration with Existing Fuzzing Platforms}
\label{sec:implementation:integration}
\toolname is designed to be platform-agnostic, leaving its integration, configuration, and use to the fuzzing platform. First, we outline two general methods for delivering network packets to network buffers within rehosting platforms. Afterward, we describe our implementation of \toolname with three representative platforms---\fuzzware, \hoedur, and \semu---each illustrating a distinct approach to binary-only rehosting and firmware fuzzing, as categorized in Table \ref{tab:rehosting-table}.

\begin{table}
\caption{Categories of binary-only rehosting approaches}
\label{tab:rehosting-table}

\begin{tabular}{ l | l } \toprule
Category  & Platforms \\
\midrule
Single-Stream & \textbf{Fuzzware}~\cite{scharnowski2022fuzzware}, AIM~\cite{feng2023aim}, P2IM~\cite{feng2020p2im}, DICE~\cite{mera2021dice} \\
Multi-Stream & \textbf{Hoedur}~\cite{scharnowski2023hoedur}, Multifuzz~\cite{chesser2024multifuzz}  \\
Spec-based & \textbf{SEmu}~\cite{zhou2022semu} \\
\bottomrule
\end{tabular}
\end{table}

\paragraph{\bfseries Integration Concepts}
Rehosting platforms must support network packet handling to enable \toolname to deliver packets to the firmware. While some firmware applications receive packets via serial interfaces like UART~\cite{romkey1988slip}, most rely on DMA-enabled peripherals to transfer packets directly into a RAM buffer. Although recent techniques can model simple DMA transfers~\cite{mera2021dice}, many network peripherals use complex DMA mechanisms that current methods do not support. Designing a generally applicable solution to this problem is inherently orthogonal to our approach; therefore, to handle it, we propose two integration approaches:

\begin{enumerate}
\item \emph{Hook-Based Integration}:
Many platforms allow debugging hooks at specific basic blocks. The emulator can inject packets into the correct buffer by placing a hook at the block where packet data is first accessed. During the execution of each hook, DMA control structures are checked to determine packet insertion points. These hooks are implemented once per MCU family.
Identifying the hook placement is an effort that needs to be taken once per HAL. For example, every STM32 MCU built with STM32's HAL accesses the received data within the same function: \texttt{HAL\_ETH\_ReadData}.

\item \emph{Peripheral Emulation}:
A more robust approach is the manual implementation of network peripherals based on MCU specifications, where the peripheral handles all MMIO interactions. This enables high-fidelity emulation and removes the need for per-sample analysis, as the emulated peripheral autonomously identifies when to insert new packets. This approach’s feasibility depends on the platform’s architecture.
\end{enumerate}

In the following, we describe how we integrated our approach with three different fuzzing frameworks.

\paragraph{\bfseries Fuzzware}
\fuzzware~\cite{scharnowski2022fuzzware} uses symbolic execution to model MMIO accesses, optimizing fuzzing input usage. Based on Unicorn~\cite{unicorn}, \fuzzware supports both integration methods. We manually implemented network peripherals for four MCUs, with each implementation averaging 435 lines of Python code.

\paragraph{\bfseries Hoedur}
\hoedur~\cite{scharnowski2023hoedur} is a multi-stream fuzzer that improves the MMIO modeling introduced by \fuzzware by introducing distinct input streams per MMIO channel for targeted mutation. Built with Rust on QEMU~\cite{bellard2005qemu}, integration with \toolname required cross-language support. We implemented hooks for packet reception and transmission for four MCUs, which detect network buffer addresses and inject packets as needed while \hoedur manages other MMIO functions. On average, each MCU implementation requires 184 lines of Rust code. In total, the integration required 823 lines of code.

\paragraph{\bfseries SEmu}
\semu~\cite{zhou2022semu} diverges from the other two platforms by using natural language processing to extract condition-action rules from MCU specifications, guiding MMIO emulation with high fidelity. Although primarily MMIO-focused, \semu includes an Ethernet peripheral for one supported MCU. We integrated \toolname using this peripheral, requiring only 34 lines of Python code.

\section{Evaluation}
\label{chap:eval}

To evaluate the effectiveness of \toolname, we conducted several experiments to determine whether and to what extent \vn{}s can improve embedded firmware fuzzing.
To this end, we implemented and tested \toolname across the three different rehosting platforms.
Furthermore, we evaluate our approach against \emnettest \cite{amusuo2023systematically}, a testing tool for embedded network stacks.
We address the following research questions in our evaluation:
\begin{itemize}
    \item \textbf{RQ1}. Is \toolname applicable to multiple fuzzing platforms?
    \item \textbf{RQ2}. What are the benefits of network protocol-aware fuzzing, and how can \toolname improve previous methods?
    \item \textbf{RQ3} How does \toolname compare to other ENS testing approaches?
    \item \textbf{RQ4}. Can \toolname improve the detection of bugs in embedded network stacks?
\end{itemize}

\begin{table*}[t]
\caption{Tested evaluation samples.}
\label{tab:sample-table}
\begin{adjustbox}{max width=\textwidth}
\begin{tabular}{llllllc}
\toprule
Vendor & MCU & Protocol Suite &  ENS & OS &Sample & Source \\
\midrule

    STM32 & nucleo-f767zi & Ethernet & LwIP & FreeRTOS & HTTP Server & \cite{gh-stm32cubef7} \\
    STM32 & nucleo-f767zi & Ethernet & NetXDuo & ThreadX & UDP Server & \cite{gh-stm32cubef7} \\ 
    STM32 & f429 & Ethernet & LwIP & Raw &TCP Echo Server  & \cite{gh-semu} \\ 
    STM32 & f429 & Ethernet & LwIP & Raw & TCP Echo Client & \cite{gh-semu} \\ 
    Nordic & nrf52840dk & BLE & nordic softdevice & zephyr & BLE Heart Rate Monitor & \cite{gh-sdk-nrf} \\ 
    Nordic & nrf52840dk & BLE & nimBLE \& nordic softdevice & nuttx & nuttx nimBLE & \cite{gh-sdk-nrf} \\ 
    Nordic & nrf52840dk & IEEE 802.15.4 & uIP & contiki-ng & Radio Ping         & \cite{gh-contiki} \\ 
    Nordic & nrf52840dk & IEEE 802.15.4 & uIP & contiki-ng & CoAP Client         & \cite{gh-contiki} \\ 
    Texas Instruments & cc2538 & IEEE 802.15.4 & uIP & contiki-ng &  SNMP Server & \cite{gh-contiki} \\ 
\bottomrule
\end{tabular}

\end{adjustbox}
\centering
\end{table*}

\subsection{Setup}

To account for the inherent randomness of the fuzzing process, we run each fuzzing campaign five times and follow the best practices for evaluating fuzzers by Klees~\etal~\cite{klees2018evaluating} and Schloegel~\etal~\cite{schloegel2024fuzzingsok}. We plotted the median and the 95\% confidence interval for the results to account for any uncertainty.

\paragraph{\bfseries Hardware Configuration}
All our experiments use the same hardware configuration: two AMD EPYC 9654 CPUs running at 3.707 Ghz (192 physical cores in total), 768 GB of RAM, and SSD memory for storage. 

\paragraph{\bfseries Platforms and Fuzzers}
We performed experiments across the three different rehosting platforms \fuzzware, \hoedur, and \semu. For each platform, we designed an ablation study, consisting of two to three configurations, to measure the impact of \toolname as precisely as possible:
\begin{itemize}
    \item Baseline (\textsc{\_base}): Each platform’s unmodified version serves as a baseline to analyze how much coverage can be attributed to the network stack
    \item Random packet (\textsc{\_rand}): Baseline extended by the ability to send network packets consisting only of random data
    \item \toolname (\textsc{\_\toolname}): Baseline with the ability to send network packets emitted by \toolname.
\end{itemize}

As \semu has the capability to send random network packets by default, we use the two configurations \semubase and \semuvn.

In addition to evaluating how \toolname enhances existing rehosting platforms, we assess its effectiveness against other network stack fuzzing tools, namely \emnettest. 
While \aflnet may appear to be a relevant baseline, fundamental differences limit its applicability in the rehosting context. Applying \aflnet to firmware would require manual source modifications (\eg to remove checksums) and a hardware-in-the-loop setup for trace collection. These requirements conflict with the fundamental principles of rehosting. Moreover, \aflnet focuses solely on the application layer and uses response codes as feedback, offering only limited insights into other protocol-layer interactions. As a result, \aflnet is not directly comparable to \toolname's broader, ENS-agnostic fuzzing capabilities.

\subsection{Sample Set}
In firmware fuzzing research, standard sample sets from prior studies~\cite{feng2020p2im,zhou2021uemu,clements2020halucinator} are used to enable a fair and consistent comparison across different works. However, existing firmware sample sets rarely include network applications, as network stacks have generally been outside the scope of past research.
To address this shortcoming, we assembled a sample set of nine applications, listed in Table~\ref{tab:sample-table}. This set includes six novel applications along with two samples from \semu~\cite{zhou2022semu} and one sample from \fuzzware~\cite{scharnowski2022fuzzware}, featuring diverse firmware across four different MCUs from three vendors, spanning six OSes and five network stacks.
These network stacks cover three of the most widely used protocol suites: 

\begin{enumerate}

\item Ethernet and TCP/IP are the foundational standards in general-purpose computing, facilitating communication across the internet. The protocol suite includes multiple protocols essential for networked communication. The samples that contain a TCP/IP stack are 1) a \textit{HTTP Server} with a static IPv4 address, 2) a \textit{UDP Server} that uses DHCP to configure its IPv4 address, 3) a \textit{TCP Echo Server}, and 4) a \textit{TCP Echo Client}. Note that the last two images were taken from the SEMU data set.

\item Bluetooth Low Energy (BLE) is a low-power Bluetooth standard optimized for close proximity communication. It operates on distinct physical channels: The advertisement channel, which is used for advertising, scanning, and connection establishment, and the data channel, which is used for actual data transmission.
For the BLE stack, the applications are 1) a \textit{BLE Heart Rate Monitor}, which is built on zephyr-os, and 2) a \textit{nimBLE} application by the nuttx operating system.

\item IEEE 802.15.4 is a radio-based protocol suite for establishing WPAN networks. It operates on the two lowest layers and is typically used with other high-level standards like 6LoWPAN or Zigbee.
The samples implementing this stack are 1) the contiki-ng \textit{Radio Ping}, which is an ICMPv6 ping application, 2) a \textit{Constrained Application Protocol (CoAP) Client} that tries to interact with a corresponding server, and 3) an \textit{SNMP Server} that features a known CVE and is part of the \fuzzware sample set.

\end{enumerate}

For \fuzzware and \hoedur, we configured firmware samples using the built-in tool to generate appropriate configurations, using provided templates wherever possible and documenting minimal manual adjustments in the configurations. For instance, certain  nRF52840~DK samples required specific MCU values in user or factory information configuration registers located in RAM. This additional memory area was manually configured for samples using this feature. All samples are published with our prototype.

\begin{table}[tb]
\caption{Setup of the \emnettest data set.}
\label{tab:emnettest-setup}
\begin{adjustbox}{max width=\columnwidth}
\begin{tabular}{ l  l  l  l l} \toprule
ENS  & CVE-ID & Type & EOS & MCU\\
\midrule
FreeRTOS-  & 2018-16523 & Div-by-zero & FreeRTOS & MPS2\\
 plus-TCP& 2018-16524 & OOB Read &  & \\
 & 2018-16526 & OOB Write &  & \\
  & 2018-16601 & Integer Underflow &  & \\
   & 2018-16603 & OOB Read & & \\
\midrule
Contiki-ng & 2021-21281 & OOB Read & Contiki-ng & TI cc2538dk \\
& 2022-26053 & OOB Write & & \\
\midrule
PicoTCP & 2020-17441 & OOB Read &FreeRTOS & STM32 F769 \\
& 2020-17442 & Integer Overflow & & \\
& 2020-17444 & Integer Overflow & & \\
& 2020-17445 & OOB Read & & \\
& 2020-24337 & Infinite Loop & & \\
\bottomrule
\end{tabular}
\end{adjustbox}
\end{table}

To enable the comparison with \emnettest, we use their dataset, which consists of twelve CVEs across three ENSs. As \emnettest tests standalone ENSs compiled for the host system rather than full firmware images, a direct coverage-based comparison is not feasible in our rehosting setup.
We used the following setup to replicate their evaluation in our rehosting environment. First, we selected three different MCUs to compile the network stacks to and integrated each ENS with an embedded OS. We then backported the vulnerabilities used by \emnettest (see Table \ref{tab:emnettest-setup}).

\subsection{Experiments}
We perform four experiments to assess \toolname's ability to improve the coverage of existing approaches and to test whether \toolname allows a fuzzer to detect bugs nested deep within the network stack.
\paragraph{\bfseries Experiment 1: \semu}
First, we performed an ablation study on \semu.
As \semu only supports a limited chipset, we selected the samples of their evaluation dataset that exhibit network behavior and performed an ablation study for these samples. We fuzzed each sample five times for 24 hours with the two versions of \semu: \semubase and \semuvn. Afterward, we collected the basic blocks each run was able to cover and evaluated them. 

\paragraph{\bfseries Experiments 2 and 3: \fuzzware \& \hoedur}
Next, we performed two similar ablation studies for \fuzzware and \hoedur.
We fuzzed each sample of the full sample set five times for 24 hours with the three different versions (e.g., \fuzzwarebase, \fuzzwarerand, and \fuzzwarevn; the same applies to \hoedur). We then again collected the basic blocks that each run was able to cover and evaluated them.

\paragraph{\bfseries Experiment 4: State-of-the-art comparison}
To compare \toolname with \emnettest, we ran \hoedurvn for 72 hours with samples containing the backported vulnerabilities and then analyzed the results. We chose \hoedur as the base rehosting platform because its usage of multiple input streams allows for the most stable emulation~\cite{scharnowski2023hoedur}.

\paragraph{\bfseries Experiment 5: Applicability beyond network stacks}
To evaluate whether \toolname's techniques generalize beyond network-based communication protocols, we extended it with support for the Modbus protocol~\cite{thomas2008modbus}, which is widely used in industrial applications. A Modbus frame consists of a device ID, a function code, a length field, a variable-length payload, and a CRC checksum. To this end, we fuzzed the heat-press firmware from the \psquaredim dataset~\cite{feng2020p2im}, which utilizes the Modbus protocol. Each configuration with \fuzzwarerand and \fuzzwarevn was run five times for 24 hours. Before starting the fuzzing campaign, we reverted previous modifications introduced by \psquaredim that had disabled the Modbus checksum verification to ensure realistic conditions.

\begin{figure*}[tb]
\begin{subfigure}[b]{0.31\textwidth}
    \includegraphics[width=\linewidth]{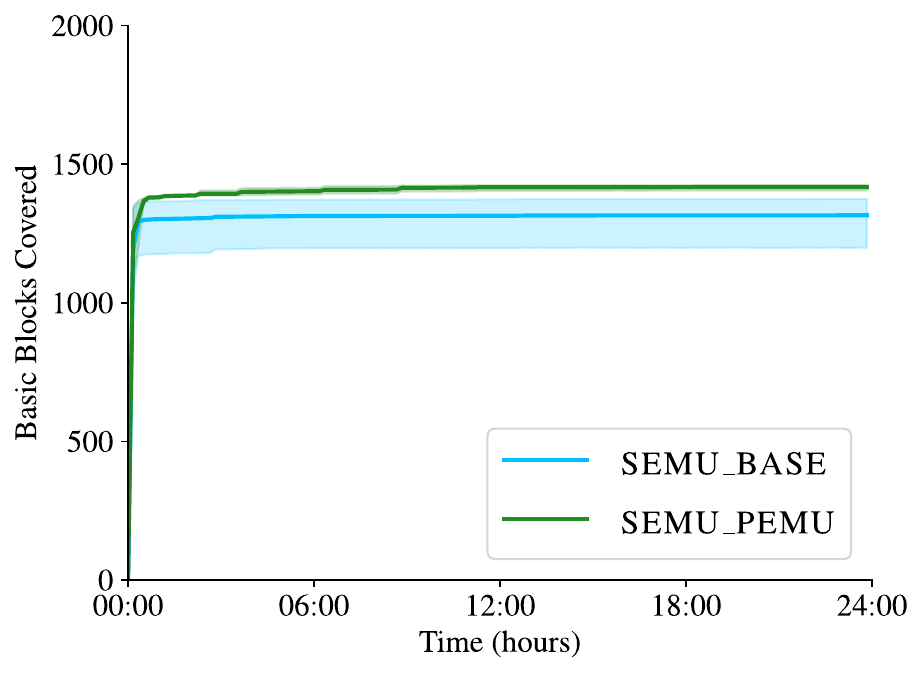}
    \caption{TCP echo server: \semu}
    \label{fig:abl_semu_tcp}
\end{subfigure}\hspace{3mm}%
\begin{subfigure}[b]{0.31\textwidth}
    \includegraphics[width=\linewidth]{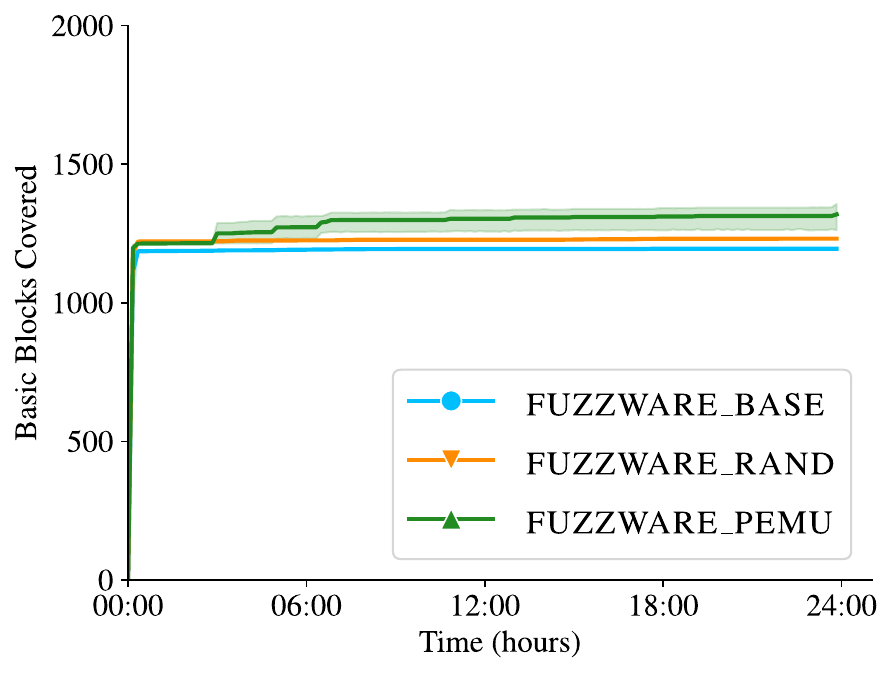}
    \caption{TCP echo server: \fuzzware}
    \label{fig:abl_fuzzware_tcp}
\end{subfigure}\hspace{3mm}%
\begin{subfigure}[b]{0.31\textwidth}
    \includegraphics[width=\linewidth]{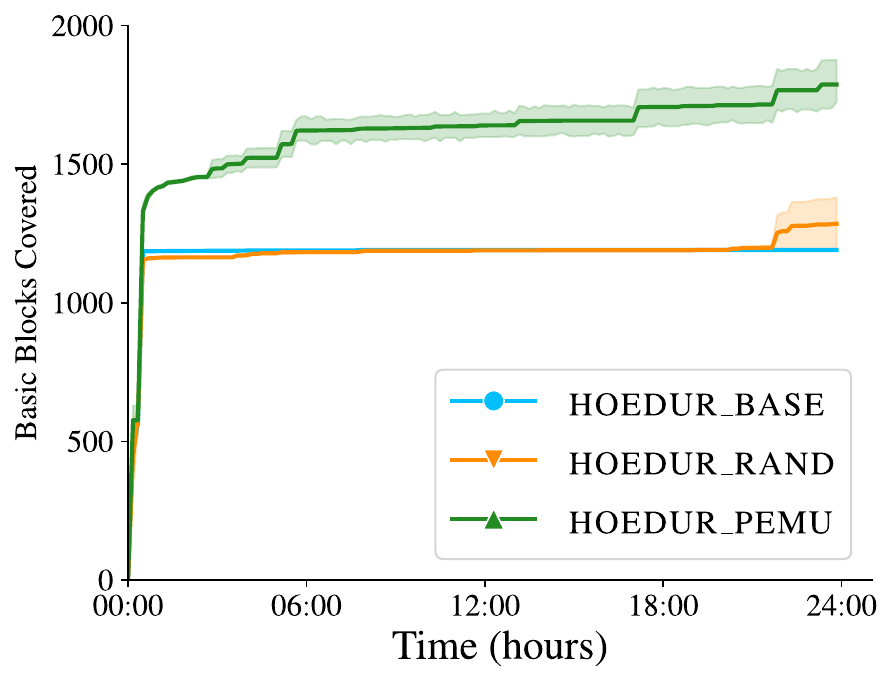}
    \caption{TCP echo server: \hoedur}
    \label{fig:abl_hoedur_tcp}
\end{subfigure}
\par\noindent\textcolor{gray}{\rule{\textwidth}{0.5pt}}\vspace{2mm}
\begin{subfigure}[b]{0.31\textwidth}
    \includegraphics[width=\linewidth]{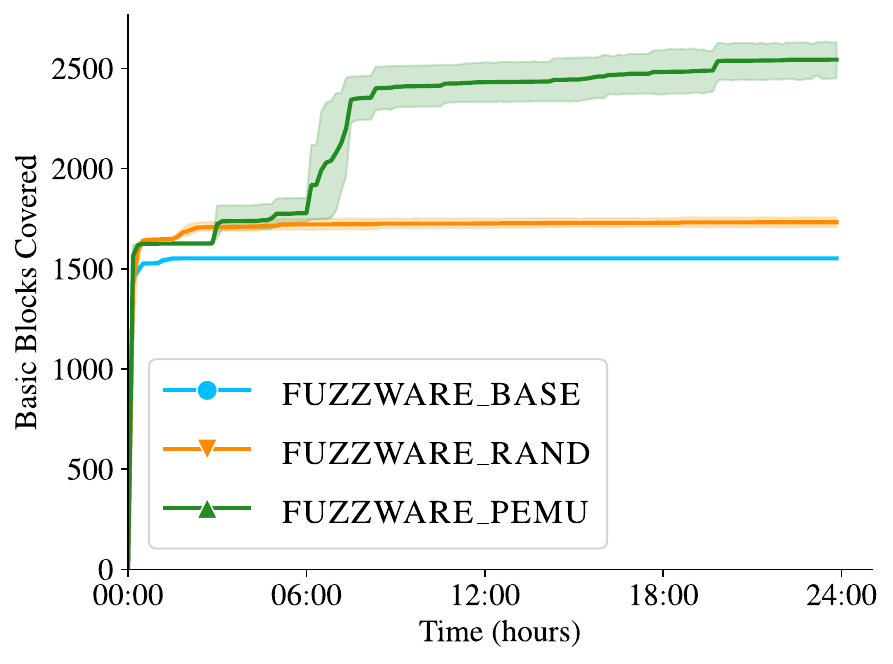}
    \caption{HTTP Server: \fuzzware}
    \label{fig:abl_fuzzware_udp}
\end{subfigure}
\rulesep
\begin{subfigure}[b]{0.31\textwidth}
    \includegraphics[width=\linewidth]{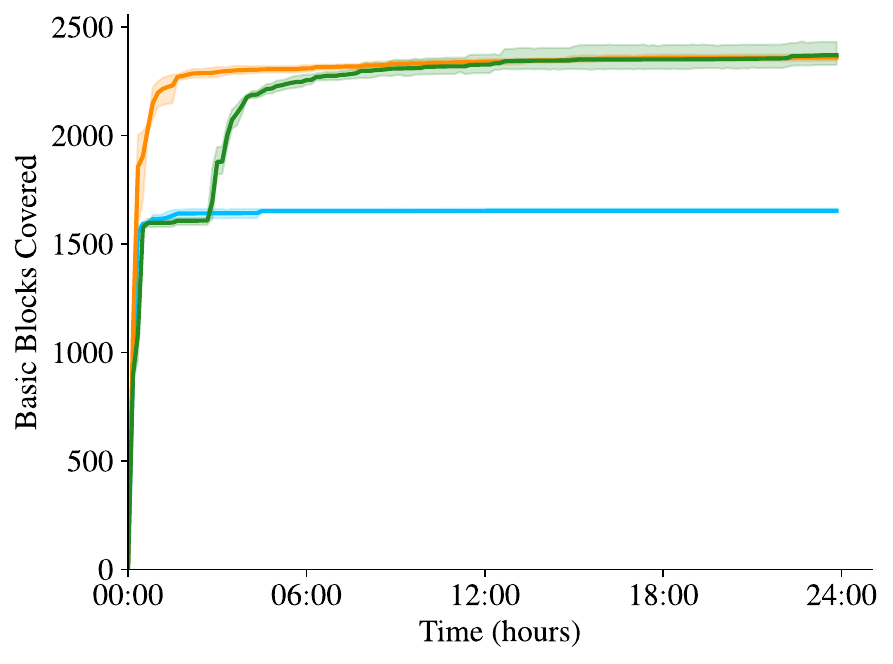}
    \caption{COAP Client: \fuzzware}
    \label{fig:abl_fuzzware_coap}
\end{subfigure}
\rulesep
\begin{subfigure}[b]{0.31\textwidth}
    \includegraphics[width=\linewidth]{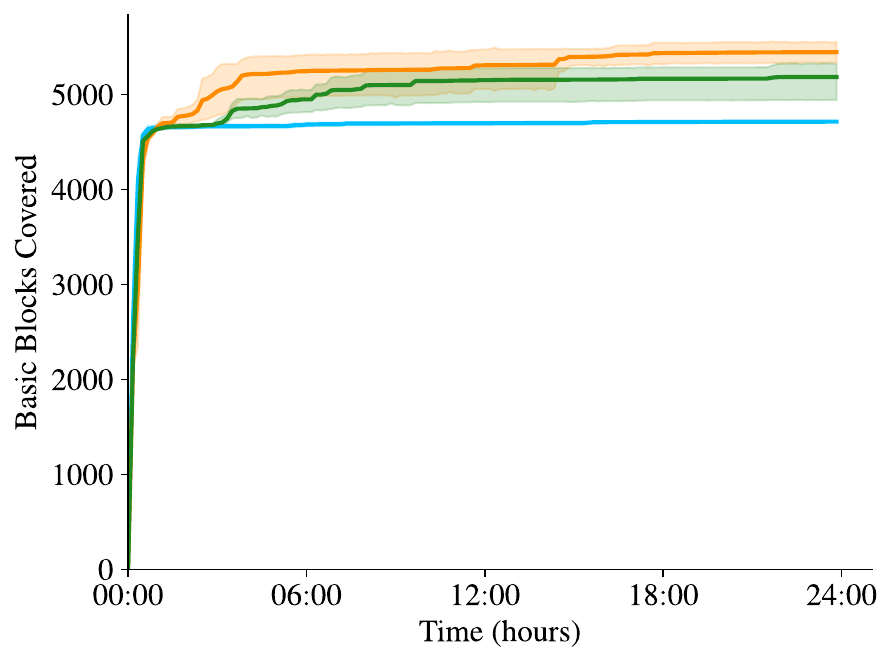}
    \caption{BLE Heart Rate Monitor: \fuzzware}
    \label{fig:abl_fuzzware_heartrate}
\end{subfigure}

\begin{subfigure}[b]{0.31\textwidth}
    \includegraphics[width=\linewidth]{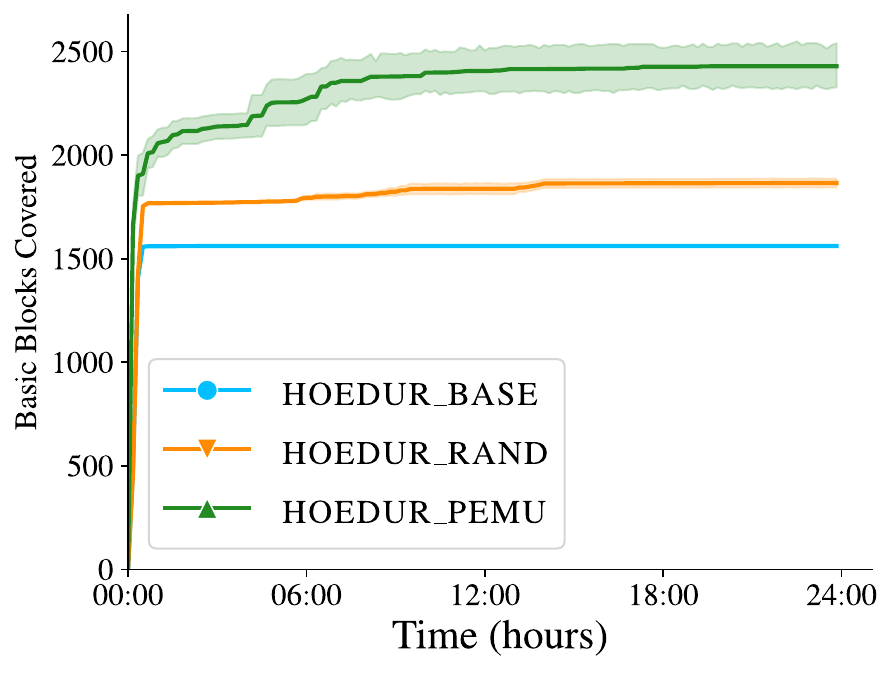}
    \caption{HTTP Server: \hoedur}
    \label{fig:abl_hoedur_udp}
\end{subfigure}\hspace{7mm}%
\rulesep
\begin{subfigure}[b]{0.31\textwidth}
    \includegraphics[width=\linewidth]{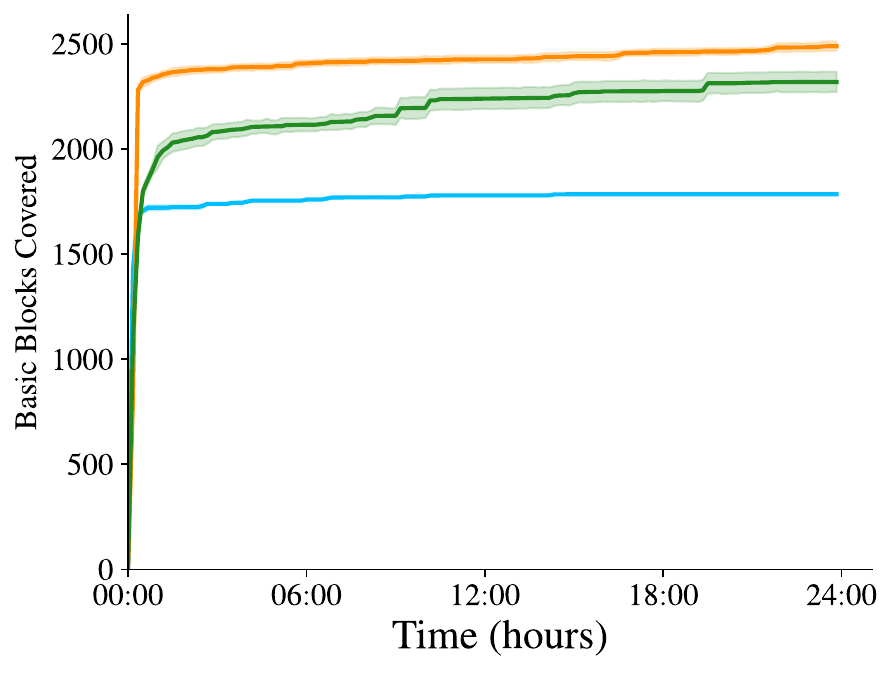}
    \caption{COAP Client: \hoedur}
    \label{fig:abl_hoedur_coap}
\end{subfigure}\hspace{6.15mm}%
\rulesep
\begin{subfigure}[b]{0.31\textwidth}
    \includegraphics[width=\linewidth]{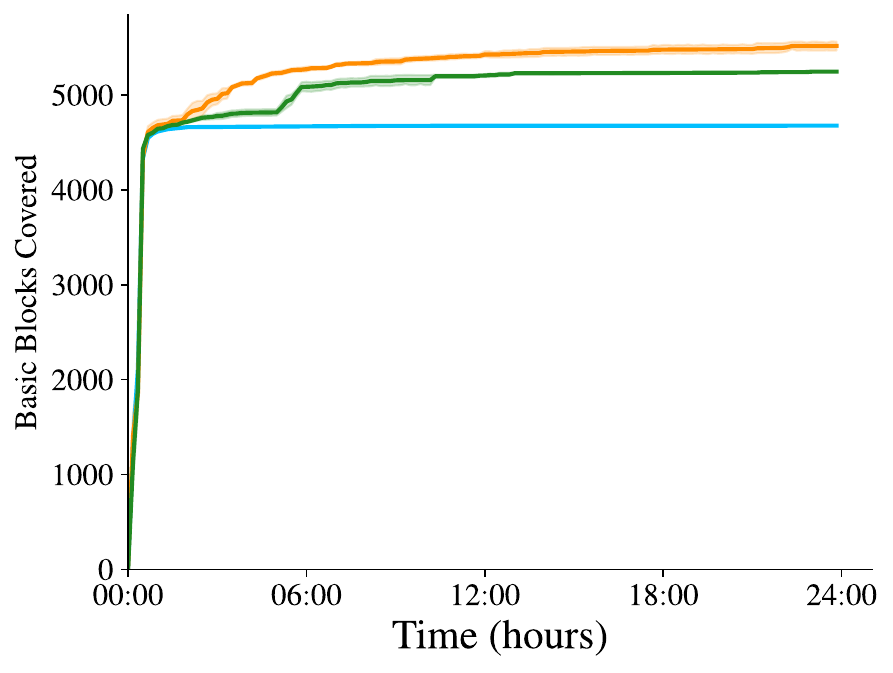}
    \caption{BLE Heart Rate Monitor: \hoedur}
    \label{fig:abl_hoedur_heartrate}
\end{subfigure}
\caption{Selected coverage plots from the different ablation studies.}
\label{fig:ablationplots}
\end{figure*}

\subsection{Results}

In the following, we summarize the experimental results in relation to the research questions. In addition to the selected plots in Figure~\ref{fig:ablationplots}, further plots can be found in the appendix. A more detailed breakdown of the results can be found in Table \ref{tab:restul-table}.

\subsubsection{RQ1}
To address the first research question on the applicability of \toolname across multiple rehosting platforms, we integrated it with three distinct platforms: \hoedur, \fuzzware, and \semu (see Section~\ref{sec:implementation}).
This demonstrates \toolname's compatibility with platforms that support (custom) network peripherals or basic block hooks and affirms the design choice of platform independence. 
As shown in Figure~\ref{fig:ablationplots}, \toolname visibly improved the coverage on all platforms despite their divergent approaches, demonstrating its versatility and applicability in embedded firmware fuzzing. 

\begin{table*}[t]
\caption[This is the short caption for the \textit{List of Tables}.]{Breakdown of the basic block coverage of each sample. \textbf{base} refers to the baseline (\semubase, \fuzzwarebase, and \hoedurbase), \textbf{rand} to the ablation configuration \fuzzwarerand and \hoedurrand, and \textbf{\toolname} to the full configuration (\semuvn, \fuzzwarevn, and \hoedurvn). The column \emph{rel imp} measures the relative improvement of \toolname over the baseline. We highlighted the results that are part of the summarized plots in Figure~\ref{fig:ablationplots}.}
\label{tab:restul-table}
\begin{adjustbox}{max width=\textwidth}
\begin{tabular}{ll | r | r r r r | r r r r | r r r r} \toprule
    & \multicolumn{1}{c}{\multirow{2}{*}{\textbf{Target}}} & \multirow{2}{*}{\textbf{\#BB in target}} & \multicolumn{4}{c}{\textbf{\#BB AVG}} & \multicolumn{4}{c}{\textbf{\#BB MAX}} & \multicolumn{4}{c}{\textbf{\#BB combined}}\\
    & &  & \textbf{base} & \textbf{rand}& \textbf{\toolname} & \emph{rel imp} & \textbf{base}& \textbf{rand}& \textbf{\toolname} & \emph{rel imp} & \textbf{base}& \textbf{rand}& \textbf{\toolname} & \emph{rel imp} \\
    \midrule

&  &  &  &  &  & & &  &  & & & & &  \\
\rowcolor{lightgray}%
\multirow{2}{*}{\rotatebox{90}{\cellcolor{white}\semu}} & TCP Echo Server & 5,212 & 1,315 & \multicolumn{1}{c}{--} & \textbf{1,417} & \textit{8\%} & 1,375 & \multicolumn{1}{c}{--} & \textbf{1,425} & \textit{4\%} & 1,375 & \multicolumn{1}{c}{--} & \textbf{1,425} & \textit{4\%} \\

& TCP Echo Client & 5,333 & 1,342 & \multicolumn{1}{c}{--} & \textbf{1,467} & \textit{9\%} & 1,406 & \multicolumn{1}{c}{--} & \textbf{1,475} & \textit{5\%} & 1,406 & \multicolumn{1}{c}{--} & \textbf{1,475} & \textit{5\%} \\
&&&&&&&&&&&&&& \\
\midrule
\rowcolor{lightgray}
\multirow{9}{*}{\cellcolor{white}\rotatebox{90}{\fuzzware}} & TCP Echo Server & 5,212 & 1,195 & 1,231 & \textbf{1,318} & \textit{10\%} & 1,196 & 1,234 & \textbf{1,374} & \textit{15\%} & 1,196 & 1,234 & \textbf{1,378} & \textit{15\%} \\
& TCP Echo Client & 5,333 & 1,227 & 1,302 & \textbf{1,661} & \textit{35\%} & 1,228 & 1,317 & \textbf{1,732} & \textit{41\%} & 1,228 & 1,327 & \textbf{1,764} & \textit{44\%} \\
    & \cellcolor{lightgray} UDP Server & \cellcolor{lightgray} 5,451 & \cellcolor{lightgray} 1,928 & \cellcolor{lightgray} 2,160 & \cellcolor{lightgray} \textbf{2,336} & \cellcolor{lightgray} \textit{21\%} & \cellcolor{lightgray} 1,938 & \cellcolor{lightgray} \textbf{2,394} & \cellcolor{lightgray} 2,375 & \cellcolor{lightgray} \textit{23\%} & \cellcolor{lightgray} 1,938 & \cellcolor{lightgray} \textbf{2,396} & \cellcolor{lightgray} 2,390 & \cellcolor{lightgray} \textit{23\%} \\
& HTTP Server & 6,579 & 1,552 & 1,732 & \textbf{2,543} & \textit{64\%} & 1,552 & 1,770 & \textbf{2,683} & \textit{73\%} & 1,552 & 1,774 & \textbf{2,705} & \textit{74\%} \\
& \cellcolor{lightgray} BLE Heart Rate Monitor & \cellcolor{lightgray} 17,469 & \cellcolor{lightgray} 4,715 & \cellcolor{lightgray} \textbf{5,446} & \cellcolor{lightgray} 5,185 & \cellcolor{lightgray} \textit{10\%} & \cellcolor{lightgray} 4,717 & \cellcolor{lightgray} \textbf{5,608} & \cellcolor{lightgray} 5,358 & \cellcolor{lightgray} \textit{14\%} & \cellcolor{lightgray} 4,723 & \cellcolor{lightgray} \textbf{5,631} & \cellcolor{lightgray} 5,380 & \cellcolor{lightgray} \textit{14\%} \\
& nuttx nimBLE & 20,260 & 6,580 & 6,588 & \textbf{6,620} & \textit{1\%} & 6,592 & 6,596 & \textbf{6,710} & \textit{2\%} & 6,602 & 6,596 & \textbf{6,711} & \textit{2\%} \\
& \cellcolor{lightgray} CoAP Client & \cellcolor{lightgray} 5,666 & \cellcolor{lightgray} 1,653 & \cellcolor{lightgray} 2,359 & \cellcolor{lightgray} \textbf{2,369} & \cellcolor{lightgray} \textit{43\%} & \cellcolor{lightgray} 1,671 & \cellcolor{lightgray} 2,382 & \cellcolor{lightgray} \textbf{2,492} & \cellcolor{lightgray} \textit{49\%} & \cellcolor{lightgray} 1,671 & \cellcolor{lightgray} 2,400 & \cellcolor{lightgray} \textbf{2,542} & \cellcolor{lightgray} \textit{52\%} \\
& SNMP Server & 4,578 & 1,445 & 2,253 & \textbf{2,267} & \textit{57\%} & 1,450 & 2,321 & \textbf{2,347} & \textit{62\%} & 1,450 & 2,354 & \textbf{2,370} & \textit{63\%} \\
& \cellcolor{lightgray} Radio Ping & \cellcolor{lightgray} 4,778 & \cellcolor{lightgray} 923 & \cellcolor{lightgray} \textbf{2,109} & \cellcolor{lightgray} 2,079 & \cellcolor{lightgray} \textit{125\%} & \cellcolor{lightgray} 929 & \cellcolor{lightgray} \textbf{2,145} & \cellcolor{lightgray} 2,108 & \cellcolor{lightgray} \textit{127\%} & \cellcolor{lightgray} 929 & \cellcolor{lightgray} \textbf{2,185} & \cellcolor{lightgray} 2,151 & \cellcolor{lightgray} \textit{132\%} \\

\midrule
\multirow{9}{*}{\rotatebox{90}{\hoedur}} & 
\cellcolor{lightgray} TCP Echo Server & 5,212 & 1,190 & 1,285 & \textbf{1,787} & \textit{50\%} & 1,198 & 1,663 & \textbf{1,915}& \textit{60\%} & 1,200 & 1,665 & \textbf{1,947}& \textit{62\%} \\
& TCP Echo Client & 5,333 & 1,218 & 1,242 & \textbf{1,828} & \textit{50\%} & 1,222 & 1,296 & \textbf{1,957}& \textit{60\%} & 1,227 & 1,299 & \textbf{2,049}& \textit{67\%} \\
& \cellcolor{lightgray}  UDP Server & \cellcolor{lightgray} 5,451 & \cellcolor{lightgray} 1,945 & \cellcolor{lightgray} 2,390 & \cellcolor{lightgray} \textbf{2,655} & \cellcolor{lightgray} \textit{37\%} & \cellcolor{lightgray} 1,946 & \cellcolor{lightgray} 2,431 & \cellcolor{lightgray} \textbf{2,667}& \cellcolor{lightgray} \textit{37\%} & \cellcolor{lightgray} 1,946 & \cellcolor{lightgray} 2,460 & \cellcolor{lightgray} \textbf{2,672} & \cellcolor{lightgray} \textit{37\%} \\
& HTTP Server & 6,579 & 1,561 & 1,865 & \textbf{2,429} & \textit{56\%} & 1,562 & 1,915 & \textbf{2,800}& \textit{79\%} & 1,562 & 1,931 & \textbf{2,802}& \textit{79\%} \\
& \cellcolor{lightgray} BLE Heart Rate Monitor & \cellcolor{lightgray} 1,7469 & \cellcolor{lightgray} 4,677 & \cellcolor{lightgray} \textbf{5,517} & \cellcolor{lightgray} 5,246 & \cellcolor{lightgray} \textit{12\%} & \cellcolor{lightgray} 4,689 & \cellcolor{lightgray} \textbf{5,635} & \cellcolor{lightgray} 5,255& \cellcolor{lightgray} \textit{12\%} & \cellcolor{lightgray} 4,689 & \cellcolor{lightgray} \textbf{5,661} & \cellcolor{lightgray} 5,319& \cellcolor{lightgray} \textit{13\%} \\
& nuttx nimBLE & 2,0260 & 6,619 & \textbf{6,627} & 6,619 & \textit{0\%} & 6,637 & \textbf{6,644} & 6,642& \textit{0\%} & 6,639 & \textbf{6,646} & 6,643& \textit{0\%} \\
& \cellcolor{lightgray} CoAP Client & \cellcolor{lightgray} 5,666 & \cellcolor{lightgray} 1,784 & \cellcolor{lightgray} \textbf{2,492} & \cellcolor{lightgray} 2,319 & \cellcolor{lightgray} \textit{30\%} & \cellcolor{lightgray} 1,786 & \cellcolor{lightgray} \textbf{2,572 } & \cellcolor{lightgray} 2,403 & \cellcolor{lightgray} \textit{35\%} & \cellcolor{lightgray} 1,786 & \cellcolor{lightgray} \textbf{2,612} & \cellcolor{lightgray} 2,435& \textit{36\%} \\
& SNMP Server & 4,578 & 1,440 & \textbf{2,519} & 2,328 & \textit{62\%} & 1,440 & \textbf{2,633} & 2,368& \textit{64\%} & 1,440 & \textbf{2,758} & 2,402& \textit{67\%} \\
& \cellcolor{lightgray} Radio Ping & \cellcolor{lightgray} 4,778 & \cellcolor{lightgray} 1,332 & \cellcolor{lightgray} \textbf{2,203} & \cellcolor{lightgray} 2,083 & \cellcolor{lightgray} \textit{56\%} & \cellcolor{lightgray} 1,332 & \cellcolor{lightgray} \textbf{2,277} & \cellcolor{lightgray} 2,115& \cellcolor{lightgray} \textit{59\%} & \cellcolor{lightgray} 1,332 & \cellcolor{lightgray} \textbf{2,349} & \cellcolor{lightgray} 2,149 & \cellcolor{lightgray} \textit{61\%} \\

\bottomrule

\end{tabular}

\end{adjustbox}
\centering
\end{table*}

\subsubsection{RQ2 -- \semu}
The results in Figure~\ref{fig:abl_semu_tcp} (and Figure~\ref{fig:semuplots} in the appendix) display the coverage achieved by \semuvn (\ie \semu with \toolname) and the coverage by \semubase (\ie \semu without any modifications).
The coverage achieved by \semuvn outperforms the baseline \semubase by 8.5\% on average. 
Due to \semu's internal architecture, it cannot send more than one packet per emulation run. Still, this improvement highlights the benefits of network-aware fuzzing. 

\subsubsection{RQ2 -- \fuzzware \& \hoedur}
\label{sec:results:fw}
Next, we discuss the results for \fuzzware and \hoedur---which both support the entire sample set---together due to the similarity of their results.
On average, \toolname is able to improve the coverage of the \fuzzware baseline by 40.7\%, while it improves the baseline of \hoedur by 39.2\%.
To highlight different aspects, we further analyze the results by protocol stack:

\paragraph{\bfseries TCP/IP}
For the four samples that implement a TCP/IP network stack (\ie \textit{HTTP Server}, \textit{UDP Server}, \textit{TCP Echo Server}, and \textit{TCP Echo Client}), the results in Table \ref{tab:restul-table} show a consistent pattern: 
For both \fuzzware and \hoedur, the ablations with \toolname clearly outperform their respective baseline by achieving 32.5\% more average coverage for \fuzzware and 48.2\% for \hoedur. When looking at the combined number of basic blocks covered over the five fuzzing runs, \toolname even improves basic-block coverage by 39\% and 61.3\%, respectively.
Notably, the second ablation (\fuzzwarerand and \hoedurrand) surpasses the baseline in both cases but is consistently outperformed by the configurations using \toolname.
These results confirm the hypothesis that random inputs progress significantly worse in network stacks that enforce strict semantic constraints on their input.
In a TCP/IP stack, these constraints are already present on the IPv4 layer (\ie the second lowest layer), including checks for IPv4 addresses, a static protocol value, and a header checksum. 
While some implementations do not enforce the checksum, allowing randomized data to progress marginally further, most packets fail at the IPv4 address validation.
In contrast, \toolname either leverages DHCP packets to dynamically assign the firmware an IPv4 address or extracts the firmware's IPv4 address by parsing outgoing packets. Consequently, \toolname is able to progress considerably further through the network stack.

\paragraph{\bfseries IEEE 802.15.4}
The results for the IEEE 802.15.4/6LoWPAN-based samples (\ie \textit{CoAP Client}, \textit{Radio Ping}, and \textit{SNMP Server}) differ from those of the TCP/IP samples. Both \fuzzwarerand and \fuzzwarevn cover up to 100\% more basic blocks than the baseline \fuzzwarebase. However, the difference in coverage between \fuzzwarerand and \fuzzwarevn is much smaller and, in some cases, yields almost identical results.
This trend is similar for \hoedur; due to the overall better coverage of the \hoedur baseline compared to \fuzzware, the relative improvement is slightly smaller---on average 53\%. Again, the improvement gains on the second ablation \hoedurrand are less significant or slightly negative.

Generally speaking, these results reflect the differences in the lower layers of the TCP/IP and the IEEE 802.15.4 network stacks. In TCP/IP, the IPv4 layer (\ie the second lowest layer) already acts as a stringent barrier, discarding all semantically and syntactically invalid packets. 
Conversely, in the IEEE 802.15.4 layer, stringent checks like checksum validation are deferred to the transport layer (\ie the layer above IPv6).
Header compression further simplifies the validation of packets as valid, even without explicit addresses.
Consequently, random input can more easily penetrate the MAC layer, 6LoWPAN adaption layer, and IPv6 network layer.
The complexity of these layers, due to compression and fragmentation and a multitude of option headers, provides a more expansive code path for coverage than TCP/IP. As \toolname prioritizes methodical exploration of header fields, it requires additional time to achieve similar coverage. 
However, manual inspection of the results revealed that, for example, \fuzzwarerand was unable to induce coverage in the application layer (\ie SNMP and CoAP). At the same time, \fuzzwarevn consistently reached the respective handlers.

For \hoedur, the difference between \hoedurrand and \hoedurvn is exacerbated by performance reasons owed to the cross-language setup, significantly slowing the execution speed. 
Compared to unmodified \hoedur the performance overhead ranges from a factor of 2 to 10. This overhead can vary depending on how many packets \toolname crafts per execution. As input lengths grow over time, execution duration increases accordingly, leading to a higher per-execution overhead of \toolname. 
However, to utilize the multi-stream aspect provided by \hoedur, our implementation is forced to request input from \hoedur in small chunks. As the fuzzing input can only be supplied within the Rust context of \hoedur, this necessitates crossing the language border multiple times for a single network packet. We discuss this implementation aspect further in Section~\ref{sec:performance}.

\paragraph{\bfseries Bluetooth Low Energy (BLE)}
The two samples that implement a BLE network stack (\ie \textit{nimBLE example} and \textit{BLE Heart Rate Monitor}) yield similar results to those of the IEEE 802.15.4 stack. 
Unlike the previous samples, both BLE samples include a proprietary component by the MCU vendor, which contains the complex logic of the BLE stack. This component explains the overall higher base coverage of both \fuzzware and \hoedur in Table \ref{tab:restul-table}. 
For the \textit{BLE Heart Rate Monitor}, both \fuzzwarerand and \fuzzwarevn achieve up to 15\% more coverage than \fuzzware's baseline. 
For \hoedur, this improvement is very similar, averaging 12\%.
This---compared to the other protocol suites---modest improvement reflects the complexity of the sample and the BLE protocol suite in general. 
Decreased emulation stability often limits the amount of packets delivered to the firmware to a single packet.
This restricts network-related coverage primarily to the BLE advertising channel (\ie advertising packets, scan packets, and connection packets) without establishing a connection that would enable access to the application logic of BLE.

The \textit{nuttx nimBLE} sample also poses a similar problem regarding its complexity: contrary to the previous sample, however, the lowest BLE layer only accepts frame types that contain the firmware's correct BLE address. Hence, packets with the wrong address are discarded immediately.
For \fuzzware, the emulation was only in one of the five runs stable enough for \toolname to successfully parse a packet and extract the firmware's BLE address. This enabled  \fuzzwarevn to outperform the other \fuzzwarebase and \fuzzwarerand regarding total and average coverage (see Table~\ref{tab:restul-table}).
However, for \hoedur, this success was not replicated, resulting in nearly identical results across all three ablations.

\paragraph{\bfseries Evaluation of Analysis Timing}
To evaluate the performance of \toolname's analysis component, we measure the detection timing of new packet types (\ie protocol detection) and values (\eg addresses and nonces), using \fuzzwarevn as an example. In this setup, we introduce an initial three-hour delay before enabling \toolname to ensure that the fuzzer reaches the point where packets are actually parsed by the firmware, which is an essential prerequisite for any analysis. 
Across all runs, \toolname consistently classified each firmware's network stack correctly, enabling valid packet generation in all cases. For most targets, the plots in Figure~\ref{fig:fuzzwareplots} in the appendix visibly exhibit an increase in coverage around the three-hour mark, corresponding to the startup of \toolname. Note that all detection times reported below are relative to this initial three-hour delay. 

The median time to detect the first protocol is 47 minutes, while the median time for the \emph{last} protocol detection is 12 hours and 33 minutes. These results show that \toolname can detect protocols reasonably fast and can continuously expand the sequence of packets it sends. The delay between protocol detections is a result of our design: after each new configuration is found, the fuzzer is given a configurable window of uninterrupted fuzzing, potentially uncovering new firmware states that will feed back information into \toolname’s analysis. Taking the UDP echo firmware as an example, \toolname detected DHCP in its first analysis iteration and added two DHCP packets to its configuration. After further fuzzing, \toolname's analysis added a UDP packet to the configuration. Next, \toolname detected that the firmware also parses ARP packets and expands the configuration. At this point, \fuzzwarevn is able to fuzz the UDP echo server's logic. During subsequent analysis runs, \toolname also added an IGMP and an ICMP4 packet to the configuration, as the firmware also implements handling for these protocols. This step-by-step expansion illustrates the versatility of \toolname and shows the importance of our dynamic, feedback-driven analysis approach.

The median time for detecting the first value is 1 hour and 35 minutes. Over the course of a fuzzing campaign, \toolname detects new protocol configurations an average of 3.75 times, while new values are detected 1.4 times on average. These values are typically either (static) addresses or identifiers like the transaction ID field in DHCP.

In the following, we provide a detailed analysis of the coverage across the individual network layers, focusing specifically on the network and transport layers. We exclude the two BLE samples, as the BLE protocol stack does not comply with the standard OSI model~\cite{gomez2012overview}, making a comparison difficult. On average, \fuzzwarevn reaches the target's \emph{network layer} 25 minutes after the three-hour delay, or 14 minutes after detecting the first protocol configuration. 
Overall, \fuzzwarevn successfully covers the network layer of all seven non-BLE samples. For the \emph{transport layer}, the average time to first coverage is 3 hours and 5 minutes after discovering the initial protocol configuration. Reaching the transport layer usually coincides with a successful detection of the target's IP address, unless the target uses a dynamic IP address. Then, \toolname assigns this address by sending the appropriate DHCP packets. \fuzzwarevn reaches the transport layer for all cases where the sample has enabled at least one transport layer protocol. The only exception is the Radio Ping sample, which does not implement any functionality above the network layer. Comparing this to our baseline, \fuzzwarerand, we find it can reach the transport layer in only a single case, and only after requiring an average of 10 hours and 21 minutes.

\subsubsection{RQ3 Comparison with \emnettest}
We ran \hoedurvn for 72 hours on the \emnettest dataset and analyzed the results. Our evaluation showed that \toolname successfully detected \emph{all} twelve known bugs in the data set (see Table \ref{tab:emnettest-setup}), demonstrating both its correctness and effectiveness.
All detected bugs were located on the network or transport layer, several requiring multiple subsequent packets to trigger the bug, which highlights \toolname's ability to handle protocol state and sequencing. Additionally, \toolname discovered three previously unknown bugs in the FreeRTOS-plus-TCP ENS, while \emnettest did not detect any new issues in this stack. Furthermore, even though the fuzzer ran for 72 hours, all bugs were rediscovered within the first 24 hours.

\subsubsection{RQ4 Bug Finding Ability of \toolname}
In addition to the twelve real-world bugs rediscovered throughout the \emnettest experiments, we identified five previously unknown bugs during the evaluation of \toolname.
Three of these bugs were found in the FreeRTOS-plus-TCP ENS. We are currently in the process of investigating and disclosing them to the vendor. 
The fourth bug is an out-of-bounds (OOB) write in LwIP. \toolname discovered it in one of the samples introduced in the \semu dataset. In contrast, \semu was unable to detect this bug that allows an attacker to control the target's PC by overwriting data structures behind the network buffer. Further analysis showed that the bug resulted from a misconfiguration during the creation of the sample by the \semu authors.
We discovered the fifth bug in the HAL of the STM32F767 MCU. It is an OOB read that arises due to improperly handled thread synchronization. The target needs to receive a valid TCP segment, which leads to a response from the firmware, to trigger the bug. Simultaneously, the firmware automatically sends periodic ARP broadcasts to the network. If the transmission of the response and the periodic broadcast overlap, it can trigger a race condition, which leads to an OOB read that can leak data to an attacker. We disclosed this to the vendor, who acknowledged it and published a dedicated security advisory.
This bug emphasizes the importance of holistic testing of network stacks in embedded firmware: although the bug does not reside in the network stack, triggering it requires the reception of a valid packet. Testing only the standalone network stack---like \emnettest does---cannot reveal such bugs, highlighting the advantages of approaches that enable an end-to-end analysis.

\subsubsection{Experiment 5: Applicability of \toolname beyond Network Stacks.}
\begin{figure}
    \centering
    \includegraphics[width=0.84\linewidth]{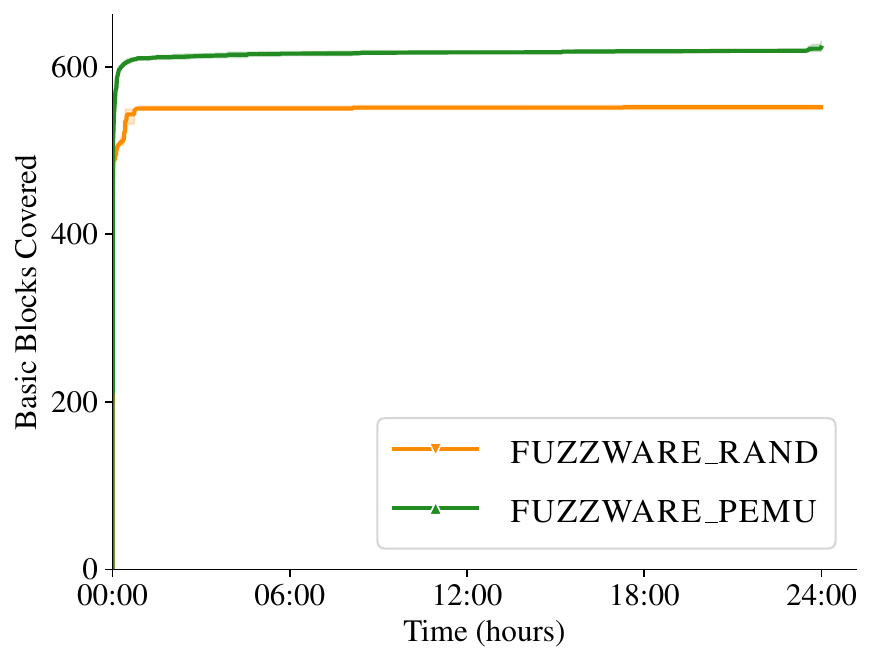}
    \caption{Coverage plot for the heat-press firmware, which utilizes the Modbus protocol.}
    \label{fig:heat-press}
\end{figure}

To further assess \toolname's applicability beyond network stacks, we fuzzed the heat-press firmware~\cite{feng2020p2im}, which communicates via the Modbus protocol. Figure~\ref{fig:heat-press} shows the coverage achieved by both ablations (\fuzzwarevn and \fuzzwarerand). As valid packets need to have a correct CRC checksum, \fuzzwarevn outperforms \fuzzwarerand by a significant margin. Manual analysis confirms that packets from \fuzzwarerand could pass the checksum verification occasionally; however, this happens too infrequently to produce meaningful new coverage. Statistically, the probability of a randomly generated packet having a correct CRC-16 checksum is very low. 
This case study demonstrates that \toolname is able to support structured communication protocols outside the network domain. 

\section{Discussion and Limitations}

In the following, we discuss potential limitations of our approach and our prototype and explore directions for future research.

\paragraph{\bfseries Low-Level Packet Transmission and DMA}%
\label{sec:requirements}

\toolname is designed to extend any firmware rehosting platform that integrates a fuzzer for firmware testing. We assume the mechanism for receiving and transmitting low-level network frames to be known, \ie we require a mechanism to control the contents of the low-level frames that the firmware receives (by writing them to the appropriate firmware memory) and a mechanism to read the frames that the firmware transmits (by reading them from firmware memory).
The main mechanism to transmit data, such as Ethernet frames or over-the-air radio frames, is DMA. If the rehosting platform handles DMA, the transfer mechanism requirement will already be fulfilled. Fully automated DMA transfer modeling is an active research area orthogonal to our work. A first step in this direction is DICE~\cite{mera2021dice}, which enables detection and rehosting of DMA streams by analyzing MMIO-based configuration patterns. The approach is geared towards simple DMA interactions typically performed by peripherals like UART, SPI, or ADC. However, complex MCUs and peripherals (like network controllers) often perform more sophisticated DMA schemes that rely on RAM-based control structures (\eg linked lists or array lists). DICE's modeling approach cannot capture these structures. Recent work by Scharnowski \etal \cite{scharnowski2025gdma} improves upon DICE, by supporting more sophisticated transfer mechanisms.
This offers the possibility to attach \toolname in an entirely generic manner.

To stay independent of the progress of automated DMA modeling research, \toolname implements alternative options to integrate with the frame transmission of the rehosting platform. One option is an available implementation of a network peripheral that facilitates the reception and transmission of low-level frames to and from firmware memory. A second option is hooking into the hardware abstraction layer (HAL) functions. Here, \toolname provides hook implementations that can be configured in the rehosting environment to be invoked when a given packet reception or transmission function is called in firmware code.

\paragraph{\bfseries Performance Impact in \hoedur}
\label{sec:performance}

As discussed in Section~\ref{sec:results:fw}, the cross-language interaction between Python and Rust,  prevalent in \hoedur, introduces a measurable performance impact. The decision to implement \toolname in Python was mainly influenced by the current ecosystem of rehosting platforms: most rehosting platforms are either written in Python or make heavy use of it~\cite{scharnowski2022fuzzware, zhou2022semu, feng2020p2im, mera2021dice, zhou2021uemu, clements2020halucinator}. 

Only recently, firmware fuzzers have started adopting Rust~\cite{scharnowski2023hoedur, chesser2024multifuzz}. 
Creating two separate implementations in Python and Rust would have been impractical, as it could have introduced inconsistencies and increased maintenance complexity. Additionally, it would have impacted comparisons across platforms.

\paragraph{\bfseries Scaling with General Rehosting Progress}

Like previous work that builds upon rehosting platforms, \toolname relies on the ability of the underlying rehosting platform to bring the firmware into a state in which it receives packets. As described in Section~\ref{sec:design:probing}, we account for this via an iterative probing approach. However, if the rehosting platform hits a roadblock unrelated to networking that prevents it from ever discovering the first layer of network functionality, the \vn of \toolname remains unused. Similarly, while \toolname provides a transparent input encapsulation mechanism, it still relies on the fuzzer to provide raw input for the application layer that is ultimately tested. As such, \toolname's benefit scales with further improvements to general rehosting platforms.

\paragraph{\bfseries Manual Effort}
The manual effort associated with using \toolname can stem from two sources. 
The first source is the integration of new protocols with \toolname. This is a one-time effort that largely depends on the protocol's complexity: structurally simple protocols like ARP, Modbus, or Ethernet can typically be added within minutes, whereas complex protocols that use compression or fragmentation may require several hours of effort. Notably, \toolname does not require exhaustive manual modeling of all protocol fields. Instead, only semantically critical fields (i.e., those essential for correct parsing) must be handled explicitly. All remaining fields can be controlled by the fuzzer. 
The second source of manual effort arises when configuring \toolname for use with a new firmware. In general, \toolname does not need any specific configuration, as it autonomously discovers new protocols and firmware-specific values like addresses and nonces. Instead, the main overhead arises when configuring the input channel through which the firmware receives its raw network frames. As discussed earlier, this effort is negligible if DMA modeling is available. Otherwise, it requires a one-time integration effort per MCU-HAL combination. First, the location of the HAL-specific abstraction function that handles the reception of network packets needs to be identified. This has to be done once per HAL as the function is consistent even across different MCUs. Second, an MCU-specific handler function---typically between 200 and 500 lines of code---that identifies the network packet buffer in RAM and injects the packet generated by \toolname needs to be written. As DMA modeling for rehosting is an actively researched field, we consider this source of manual effort a temporary limitation imposed by the current capabilities of the underlying rehosting systems, rather than an inherent constraint of \toolname.

\paragraph{\bfseries General Fuzzing Limitations}
One of the most common limitations in general-purpose fuzzing is handling encryption, as mutations applied to encrypted data are unlikely to also produce validly encrypted data. To a certain degree, \toolname can help solve this for firmware fuzzing, as encryption algorithms can be integrated into \toolname in a similar manner to compression. However, \toolname cannot handle cases where a pre-shared secret or private key is required for authentication. In this case, reverse-engineering is necessary to either extract the secret or to replace, \eg a certificate to enable \toolname to act as an authenticated server. Fortunately, there are many use cases where this is not necessary. For example, most embedded web servers do not require the client (in this case, the fuzzer + \toolname) to authenticate itself.

\section{Related Work}%
\label{sec:related_work}
Next, we discuss existing research on the two main areas on which the majority of our work focuses.

\paragraph{\bfseries Firmware Rehosting and Fuzzing}
Our work is based on the huge body of work in the field of embedded rehosting and fuzzing~\cite{feng2020p2im, mera2021dice, scharnowski2022fuzzware, scharnowski2023hoedur, clements2020halucinator, zhou2021uemu, zhou2022semu, ruge2020frankenstein,muench2018wycinwyc, zheng2019firm-afl, hernandez2022firmwire, cao2020device, seidel2023safirefuzz, chesser2024multifuzz, gao2020fuzz}.
Especially \hoedur and \fuzzware by Scharnowski et al.~\cite{scharnowski2023hoedur, scharnowski2022fuzzware} and \semu by Zhou \etal~\cite{zhou2022semu} serve as the base for the implementation of our approach. We expect that other rehosting methods can also benefit from the virtual network provided by \toolname. 
Beyond rehosting, research has also expanded to fuzzing embedded hardware~\cite{wang2024syztrust, ma2023no, kim2021pasan, aafer2021_smarttvfuzzing, chen2018iotfuzzer} or using partial emulation to solve the hardware-dependency of firmware~\cite{muench_avatar_2018, koscher2015surrogates, kammerstetter2014prospect, mera2024shift}.

\paragraph{\bfseries Network Application Fuzzing}
\label{sec:background:network-fuzz}

Network application fuzzing for general-purpose software has been extensively studied, leading to tools such as \textsc{AFLNet}~\cite{pham2020aflnet}, \textsc{SGFuzz}~\cite{ba2022sgfuzz}, \textsc{StateAFL}~\cite{natella2022stateafl}, \textsc{Bleem}~\cite{luo2023bleem}, and \textsc{Fuzztruction-Net}~\cite{bars2024fuzztructionnet}, each introducing different methods to address this challenge. Existing approaches primarily target the application layer, leveraging modified network traffic seeds or OS functionalities to construct packets that encapsulate fuzzing input.

\section{Conclusion}

In this work, we introduced \toolname, the first automated approach for fuzzing the network stacks of embedded firmware. 
We tackle the challenges of ensuring well-formed encapsulation within network packets and awareness of the protocol state.
\toolname focuses on the \emph{in-depth} exploration of firmware by dynamically analyzing its network-related behavior and adapting to it to provide relevant fuzzing inputs encapsulated in seemingly real network packets. Due to its robust and automated analysis approach, it is possible to apply \toolname without requiring domain knowledge. 
To demonstrate how existing fuzzing methods can benefit from a virtual network stack, we integrated \toolname with three state-of-the-art rehosting platforms.
In a comprehensive evaluation, we showed that our method provides an effective way of analyzing and fuzzing firmware using common types of embedded networking protocols.

\begin{acks}
We thank the anonymous reviewers and our shepherd for their valuable feedback.
This work was funded by the European Research Council (ERC) under the consolidator grant RS$^3$ (101045669) and by the Deutsche Forschungsgemeinschaft (DFG, German Research Foundation) under Germany's Excellence
Strategy (EXC 2092 CASA --- 390781972).
This material is based upon work supported by the National Science Foundation under Award No. 2232915, 2146568, 2442984, and 2247954, as well as by the Advanced Research Projects Agency for Health (ARPA-H) under Contract No. SP4701-23-C-0074. Any opinions, findings, and conclusions or recommendations expressed in this material are those of the author(s) and do not necessarily reflect the views of the NSF or ARPA-H.
\end{acks}

\bibliographystyle{plain}
\balance
\bibliography{z_strings, z_references, z_autogenerated}

\appendix
\renewcommand{\thesection}{\Alph{section}}

\section*{A \semu Results}
\begin{figure}[b]
\centering
\begin{adjustbox}{width=0.6\textwidth}
\begin{minipage}{\textwidth}
\begin{subfigure}{0.48\textwidth}
    \includegraphics[width=\linewidth]{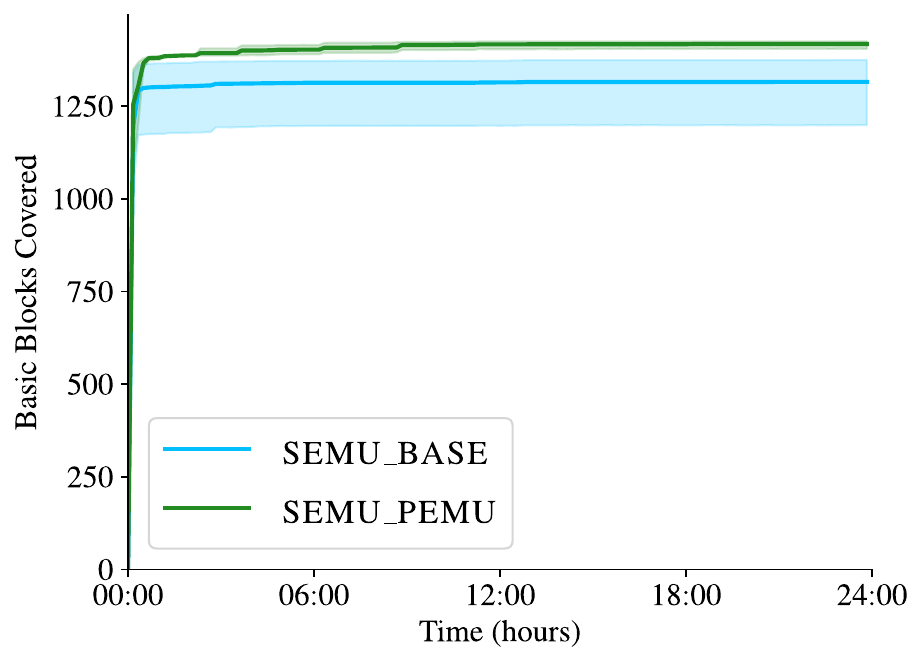}
    \caption{f429 TCP 
    Echo Server}
    \label{fig:semu_tcp_server}
\end{subfigure} 

\begin{subfigure}{0.48\textwidth}
    \includegraphics[width=\linewidth]{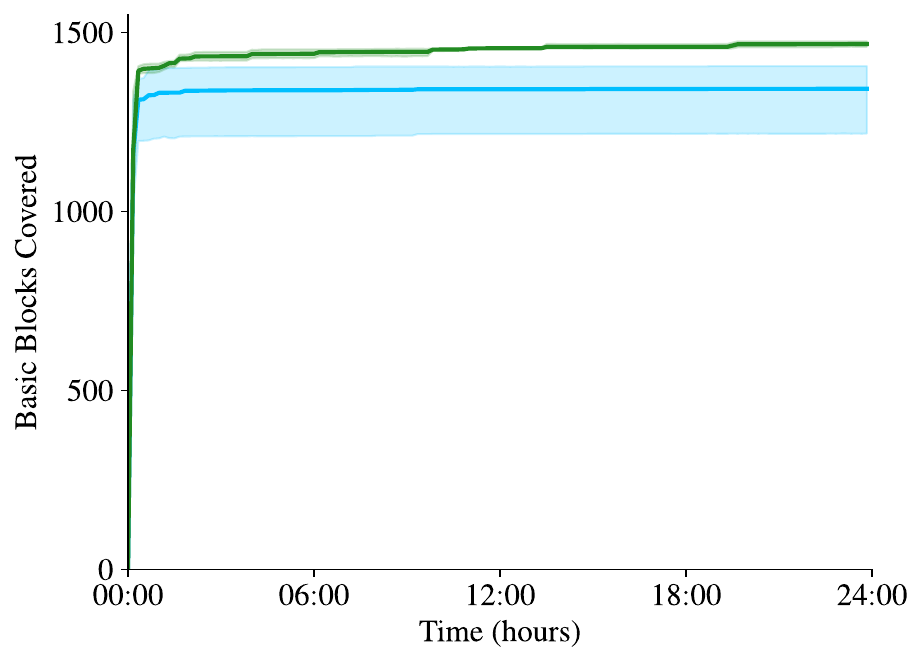}
    \caption{f429 TCP Echo Client}
    \label{fig:semu_tcp_client}
\end{subfigure}
\end{minipage}
\end{adjustbox}
\caption{Coverage plots of the ablation study on \semu}
\label{fig:semuplots}
\end{figure}

We present the results of the \semu experiment in Figure~\ref{fig:semuplots}. The natural language processing approach by \semu only supports a limited data set Therefore, we were only able to evaluate two samples.

\newpage

\section*{B \fuzzware and \hoedur Results}
\begin{figure}[h!]
\begin{adjustbox}{width=0.48\textwidth} 
\begin{minipage}{\textwidth}
\begin{subfigure}[b]{0.48\textwidth}
    \includegraphics[width=\linewidth]{figures/eval/Fuzzware/Figure_static_ip.pdf}
    \caption{f767 HTTP Server}
    \label{fig:http_server}
\end{subfigure}
\begin{subfigure}[b]{0.48\textwidth}
    \includegraphics[width=\linewidth]{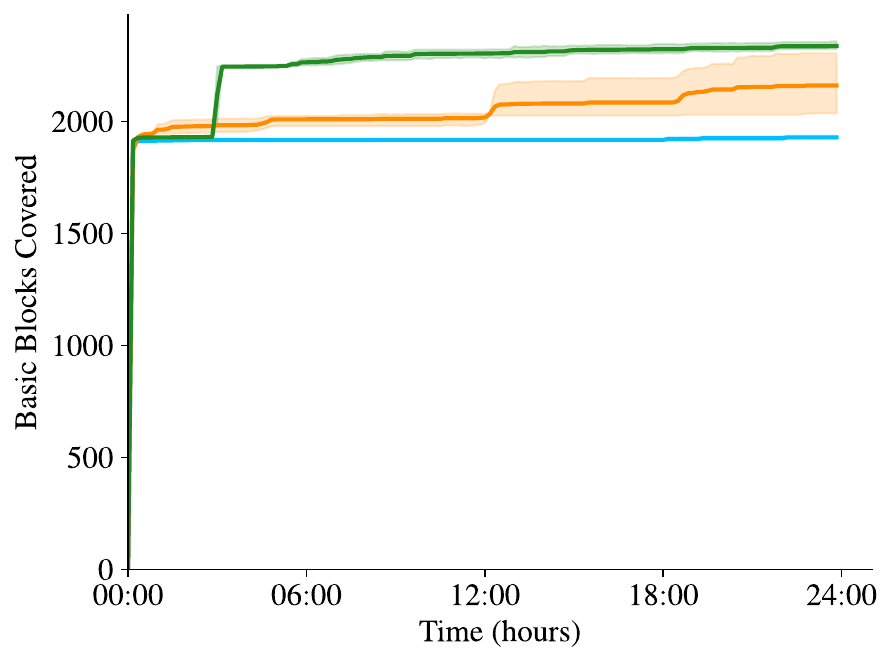}
    \caption{f767 UDP Server}
    \label{fig:udp_server}
\end{subfigure}

\begin{subfigure}[b]{0.48\textwidth}
    \includegraphics[width=\linewidth]{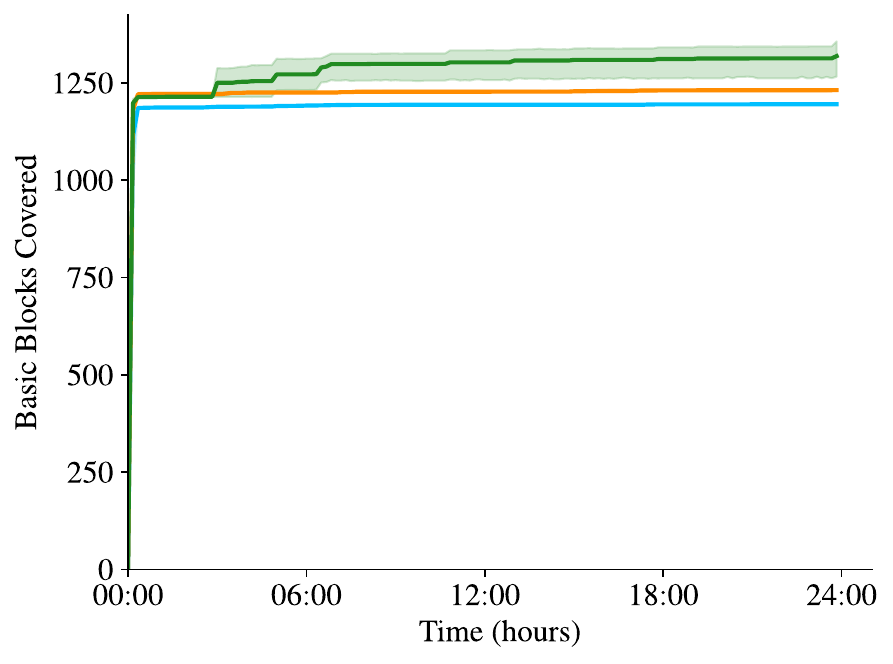}
    \caption{f429 TCP Echo Server}
    \label{fig:fuzzware_tcp_server}
\end{subfigure}
\begin{subfigure}[b]{0.48\textwidth}
    \includegraphics[width=\linewidth]{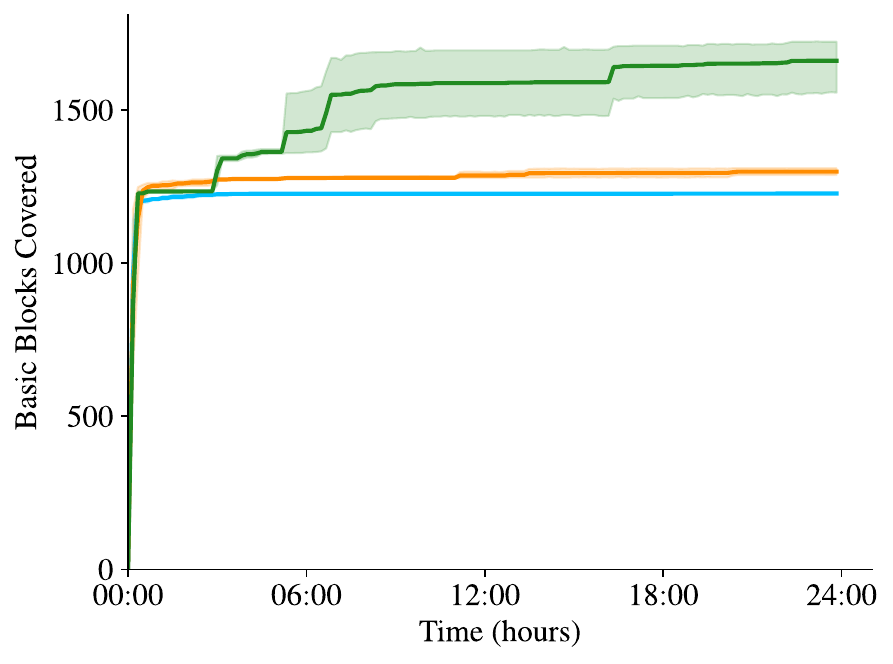}
    \caption{f429 TCP Echo Client}
    \label{fig:fuzzware_tcp_client}
\end{subfigure}

\begin{subfigure}[b]{0.48\textwidth}
    \includegraphics[width=\linewidth]{figures/eval/Fuzzware/Figure_nrf52840_central_hr.pdf}
    \caption{nrf52840dk BLE Heart Rate Monitor}
    \label{fig:heart_rate}
\end{subfigure}
\begin{subfigure}[b]{0.48\textwidth}
    \includegraphics[width=\linewidth]{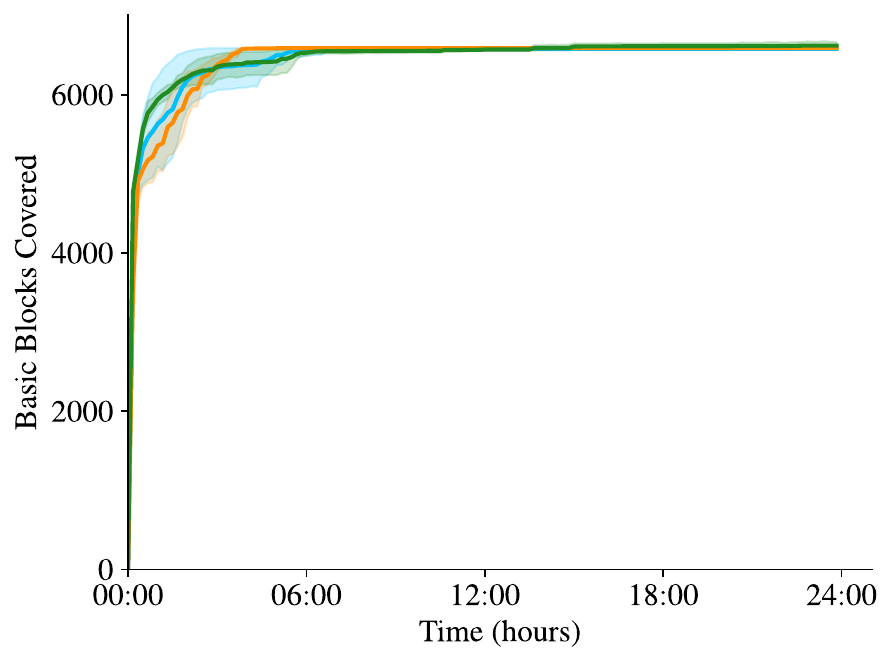}
    \caption{nrf52840dk nuttx nimBLE}
    \label{fig:nimble}
\end{subfigure}

\begin{subfigure}[b]{0.48\textwidth}
    \includegraphics[width=\linewidth]{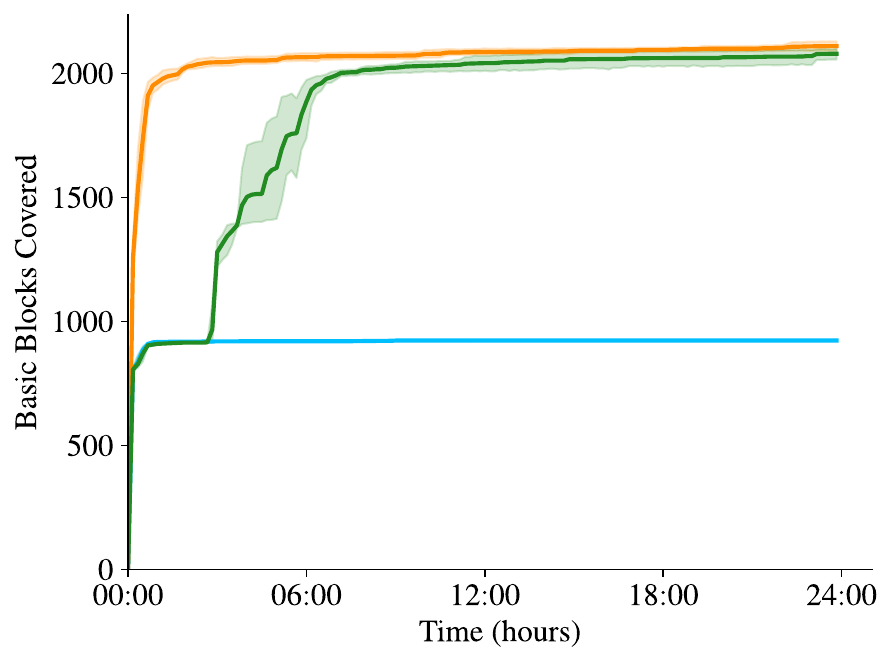}
    \caption{nrf52840dk Radio Ping}
    \label{fig:hello_world}
\end{subfigure}%
\begin{subfigure}[b]{0.48\textwidth}
    \includegraphics[width=\linewidth]{figures/eval/Fuzzware/Figure_nrf52840_coap-client.pdf}
    \caption{nrf52840dk CoAP Client}
    \label{fig:coap_client}
\end{subfigure}%

\begin{subfigure}[b]{0.48\textwidth}
    \includegraphics[width=\linewidth]{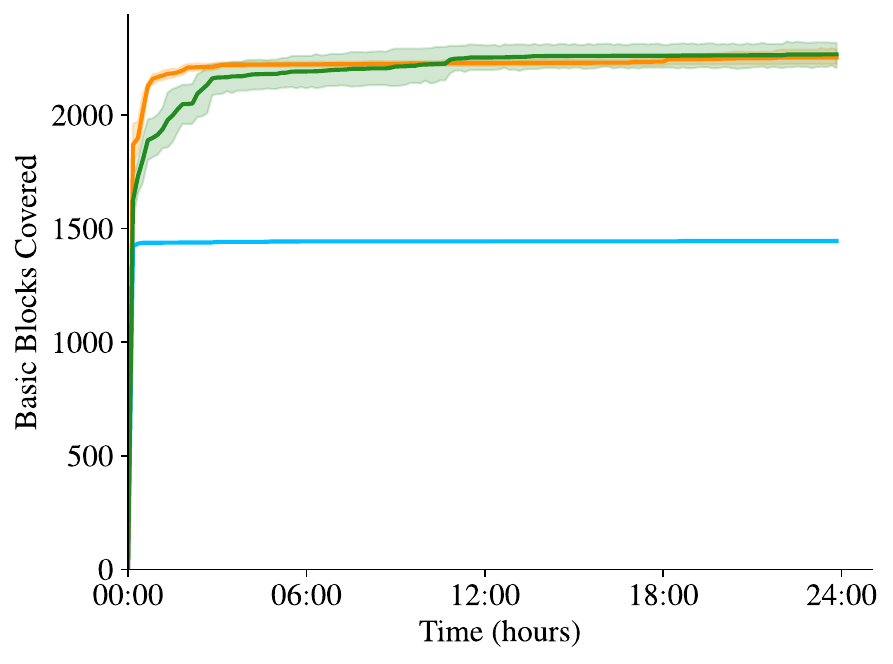}
    \caption{cc2538 SNMP Server}
    \label{fig:snmp_server}
\end{subfigure}

\end{minipage}
\end{adjustbox}
\caption{Coverage plots of the ablation study on \fuzzware}
\label{fig:fuzzwareplots}
\end{figure}

\begin{figure*}[h]
\begin{adjustbox}{width=0.65\textwidth} 
\begin{minipage}{\textwidth}
\centering
\begin{subfigure}[b]{0.3\textwidth}
    \includegraphics[width=\linewidth]{figures/eval/Hoedur/Figure_stm32-f7xx-http_static-.svg.pdf}
    \caption{f767 HTTP Server}
    \label{fig:hoedur_http_server}
\end{subfigure}
\begin{subfigure}[b]{0.3\textwidth}
    \includegraphics[width=\linewidth]{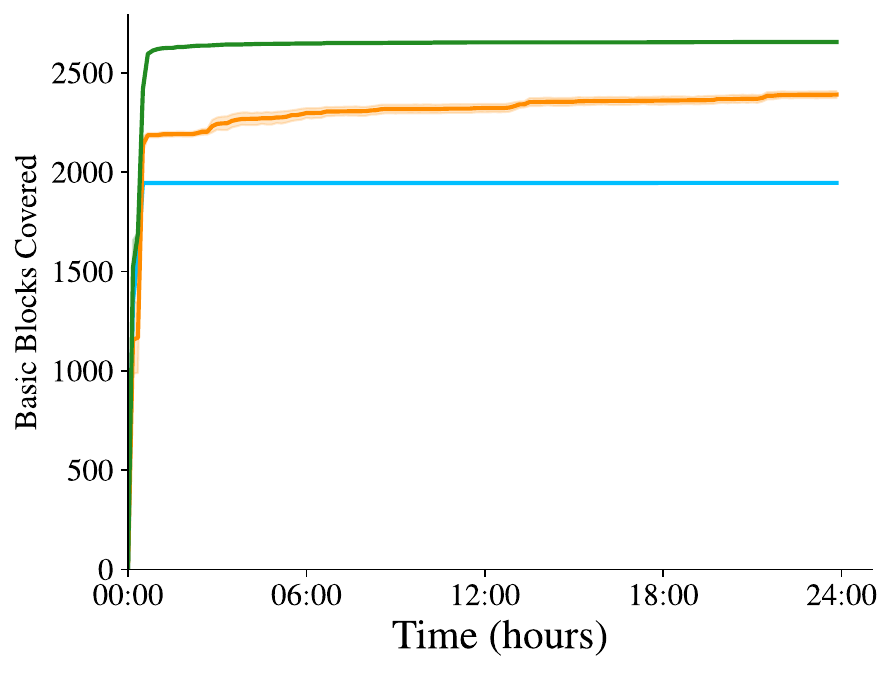}
    \caption{f767 UDP Server}
    \label{fig:hoedur_udp_server}
\end{subfigure}
\begin{subfigure}[b]{0.3\textwidth}
    \includegraphics[width=\linewidth]{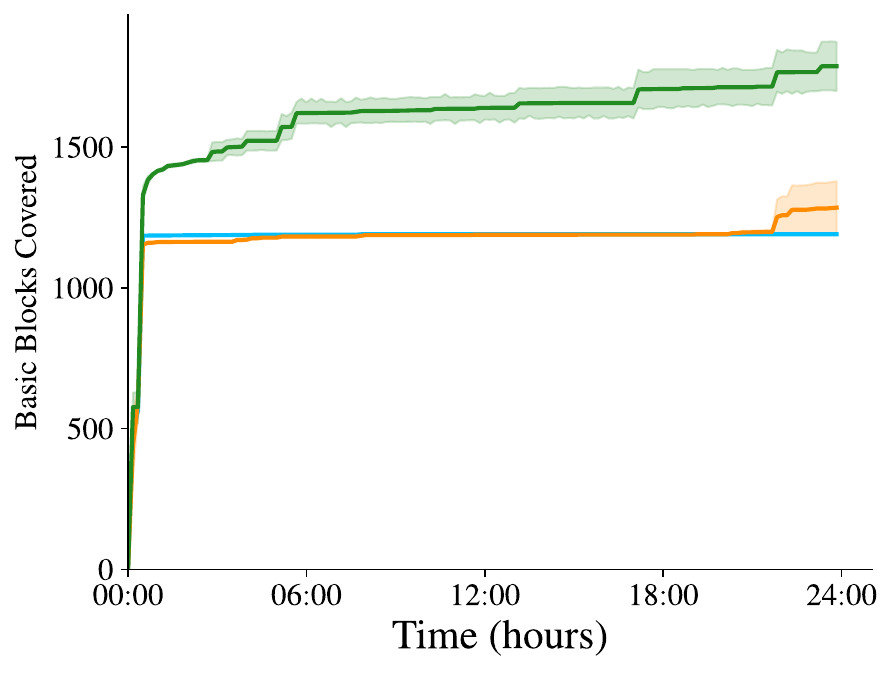}
    \caption{f429 TCP Echo Server}
    \label{fig:hoedur_fuzzware_tcp_server}
\end{subfigure}
\begin{subfigure}[b]{0.3\textwidth}
    \includegraphics[width=\linewidth]{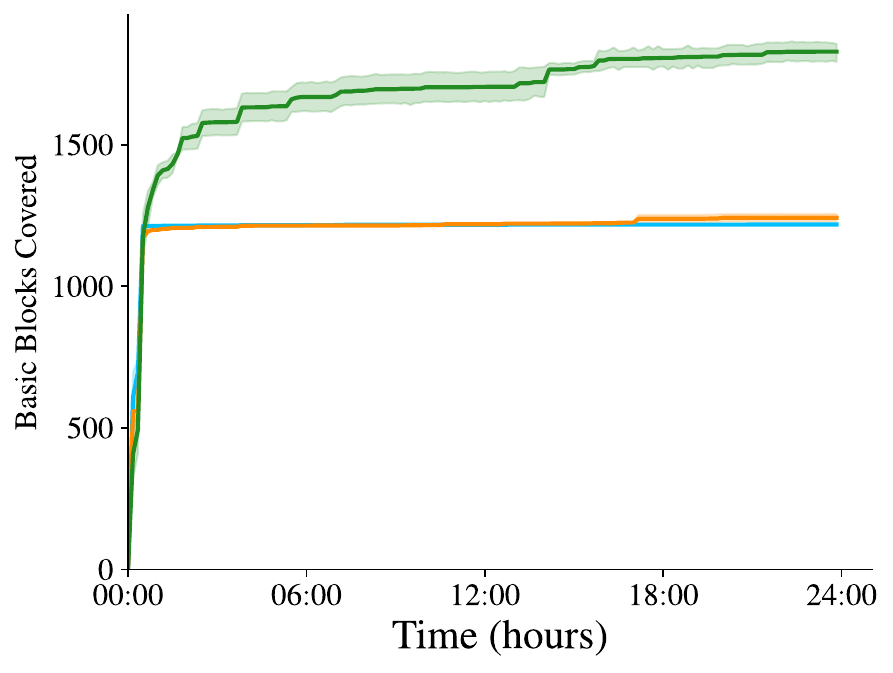}
    \caption{f429 TCP Echo Client}
    \label{fig:hoedur_fuzzware_tcp_client}
\end{subfigure}
\begin{subfigure}[b]{0.3\textwidth}
    \includegraphics[width=\linewidth]{figures/eval/Hoedur/Figure_nordic-nrf52840_central_hr-.svg.pdf}
    \caption{nrf52840dk BLE Heart Rate Monitor}
    \label{fig:hoedur_heart_rate}
\end{subfigure}
\begin{subfigure}[b]{0.3\textwidth}
    \includegraphics[width=\linewidth]{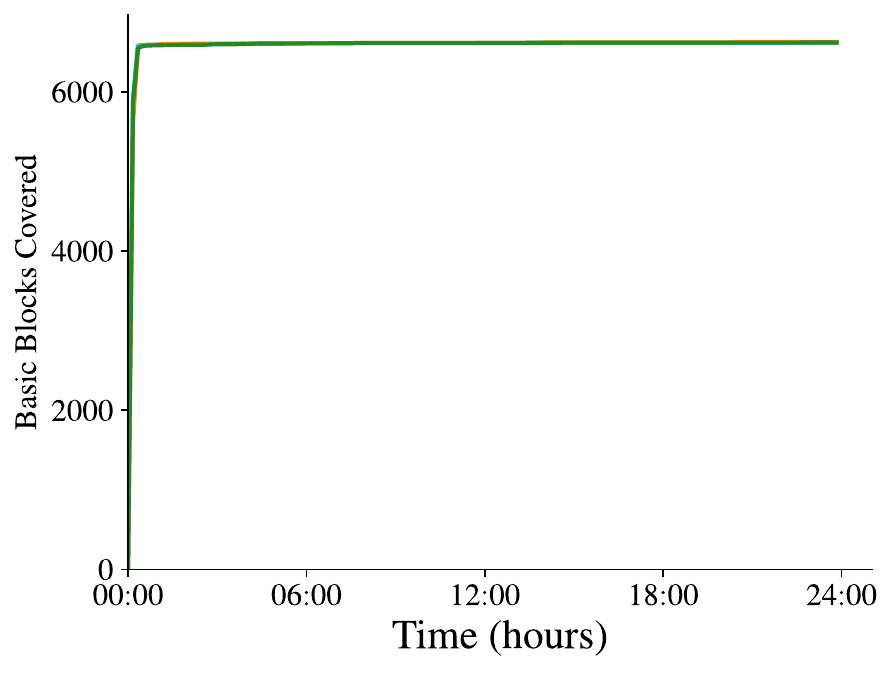}
    \caption{nrf52840dk nuttx nimBLE}
    \label{fig:hoedur_nimble}
\end{subfigure}
\begin{subfigure}[b]{0.3\textwidth}
    \includegraphics[width=\linewidth]{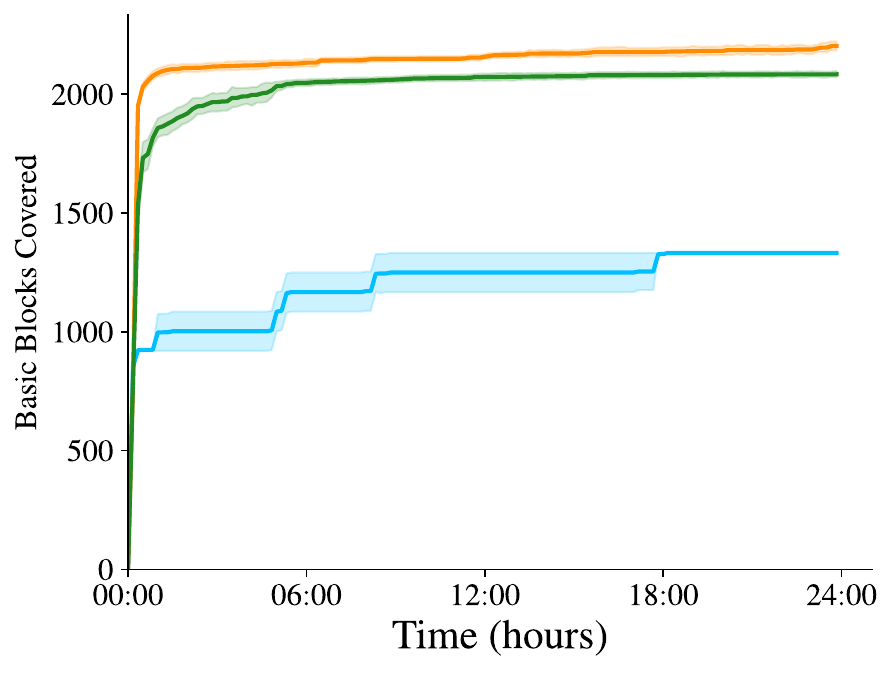}
    \caption{nrf52840dk Radio Ping}
    \label{fig:hoedur_hello_world}
\end{subfigure}%
\begin{subfigure}[b]{0.3\textwidth}
    \includegraphics[width=\linewidth]{figures/eval/Hoedur/Figure_nordic-nrf52840_coap-client-.svg.pdf}
    \caption{nrf52840dk CoAP Client}
    \label{fig:hoedur_coap_client}
\end{subfigure}%
\begin{subfigure}[b]{0.3\textwidth}
    \includegraphics[width=\linewidth]{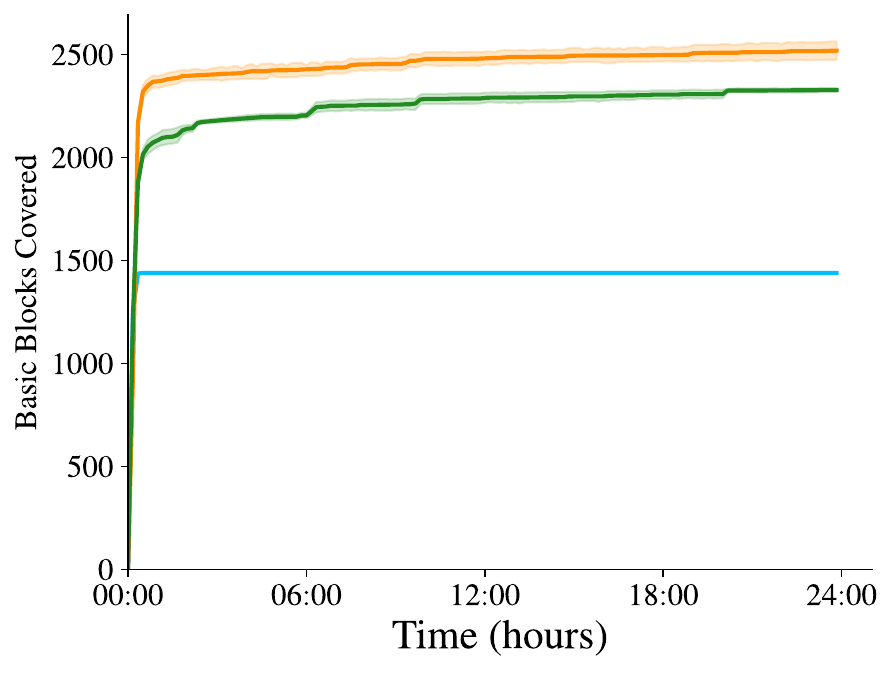}
    \caption{cc2538 SNMP Server}
    \label{fig:hoedur_snmp_server}
\end{subfigure}
\caption{Coverage plots of the ablation study on \hoedur}
\label{fig:hoedurplots}
\end{minipage}
\end{adjustbox}
\end{figure*}

The collected results of the evaluation with \fuzzware and \hoedur are displayed in Figure~\ref{fig:fuzzwareplots} and Figure~\ref{fig:hoedurplots}. Both fuzzing platforms support the entire dataset. The plots show the median and the 95\% confidence interval.


\end{document}